\documentclass{nature}

\usepackage[numbers,numbers,super,comma,sort&compress]{natbib}
\makeatletter
\def\@mb@citenamelist{cite,citep,citet,citealp,citealt,citepalias,citetalias}
\makeatother
\usepackage{multibib}
\newcites{sec}{Other references}

\usepackage{chemformula} 
\usepackage{physics} 
\usepackage{graphicx}
\usepackage{dcolumn}
\usepackage{bm}
\usepackage{chemformula} 
\usepackage{times} 
\usepackage{tabularx}
\usepackage{physics}
\usepackage{xcolor}

\usepackage{setspace}
\usepackage{amsfonts}
\usepackage{amssymb}
\usepackage{amsmath}

\usepackage{float}
\usepackage{placeins}
\usepackage{graphicx}

\usepackage[export]{adjustbox}

\usepackage[nomarkers,nolists,figuresonly]{endfloat}

\newcounter{mypostfigure} 

\AtBeginFigures{\setcounter{mypostfigure}{0}} 
\AtBeginFigures{\setcounter{mypostfigure}{0}} 

\usepackage[normal,normal,bf,labelformat=empty]{caption}

\newcounter{mybodyfigure}
\newcounter{myedfigure}

\newcommand{\beginbodyfigures}{\renewcommand{\thefigure}{{\themybodyfigure}}}

\newcommand{\beginedfigures}{\renewcommand{\thefigure}{{\themyedfigure}}}

\newcommand{\stepbodyfigure}{\refstepcounter{mybodyfigure}}

\DeclareRobustCommand{\bodyfigure}[1]{\stepbodyfigure\label{#1}{\themybodyfigure}}

\newcommand{\bodyfigurelabel}[1]{\bf{Figure \bodyfigure{#1}:}}

\makeatletter
\g@addto@macro\caption@prepareslc{%
  \renewcommand{\stepbodyfigure}{\caption@l@stepcounter{mybodyfigure}}}
\makeatother

\newcommand{\stepedfigure}{\refstepcounter{myedfigure}}

\DeclareRobustCommand{\edfigure}[1]{\stepedfigure\label{#1}{\themyedfigure}}

\newcommand{\edfigurelabel}[1]{\bf{Extended Data Figure \edfigure{#1}:}}

\makeatletter
\g@addto@macro\caption@prepareslc{%
  \renewcommand{\stepedfigure}{\caption@l@stepcounter{myedfigure}}}
\makeatother

\usepackage{hyperref}
\usepackage{verbatim}

\usepackage{ulem} 
\usepackage{rotating} 
\usepackage{soul}

\usepackage{color} 

\usepackage[colorinlistoftodos]{todonotes}

\makeatletter
\newsavebox\myboxA
\newsavebox\myboxB
\newlength\mylenA

\newcommand*\xoverline[2][0.75]{%
    \sbox{\myboxA}{$\m@th#2$}%
    \setbox\myboxB\null
    \ht\myboxB=\ht\myboxA%
    \dp\myboxB=\dp\myboxA%
    \wd\myboxB=#1\wd\myboxA
    \sbox\myboxB{$\m@th\overline{\copy\myboxB}$}
    \setlength\mylenA{\the\wd\myboxA}
    \addtolength\mylenA{-\the\wd\myboxB}%
    \ifdim\wd\myboxB<\wd\myboxA%
       \rlap{\hskip 0.5\mylenA\usebox\myboxB}{\usebox\myboxA}%
    \else
        \hskip -0.5\mylenA\rlap{\usebox\myboxA}{\hskip 0.5\mylenA\usebox\myboxB}%
    \fi}
\makeatother

\title{Fragile superconductivity in a Dirac metal} 

\author{Chris J. Lygouras$^{1}$, Junyi Zhang$^{1}$, Jonah Gautreau$^{2}$, Mathew Pula$^{2}$, Sudarshan Sharma$^{2}$, Shiyuan Gao$^{1,3}$, Tanya Berry$^{1,3}$, Thomas Halloran$^{1}$, Peter Orban$^{1}$, Gael Grissonnanche$^{4,5}$, Juan R. Chamorro$^{1,3}$, Kagetora Mikuri$^{6}$, Dilip K. Bhoi$^{6}$, Maxime A. Siegler$^{3}$, Kenneth K. Livi$^{13}$, Yoshiya Uwatoko$^{6}$, Satoru Nakatsuji$^{1,6,7,8,9,10}$, B. J. Ramshaw$^{4,10}$, Yi Li$^{1}$, Graeme M. Luke$^{2,11}$, Collin L. Broholm$^{1,12,13}$, and Tyrel M. McQueen$^{1,3,13}$} 

\begin{document}

\maketitle

\begin{affiliations} 
\item Institute for Quantum Matter and William H. Miller III Department of Physics and Astronomy, Johns Hopkins University, Baltimore, Maryland 21218, USA
\item Department of Physics and Astronomy, McMaster University,  Hamilton, Ontario, L8S 4M1, Canada 
\item Department of Chemistry, Johns Hopkins University, Baltimore, Maryland, 21218, USA 
\item Laboratory of Atomic and Solid State Physics, Cornell University, Ithaca, NY, USA
\item Kavli Institute at Cornell for Nanoscale Science, Ithaca, NY, USA
\item Institute for Solid State Physics (ISSP), University of Tokyo, Kashiwa, Chiba, 277-8581, Japan 
\item Department of Physics, University of Tokyo, Bunkyo-ku, Tokyo 113-0033, Japan
\item Trans-scale Quantum Science Institute, University of Tokyo, Bunkyo-ku, Tokyo 113-8654, Japan
\item CREST, Japan Science and Technology Agency (JST), 4-1-8 Honcho Kawaguchi, Saitama, 332-0012, Japan 
\item Canadian Institute for Advanced Research, Toronto, M5G 1Z7, ON, Canada 
\item TRIUMF, Vancouver, British Columbia, V6T 2A3, Canada 
\item NIST Center for Neutron Research, National Institute of Standards and Technology, Gaithersburg, Maryland, 20899, USA 
\item Department of Materials Science and Engineering, Johns Hopkins University, Baltimore, Maryland, 21218, USA 
\end{affiliations}

\begin{abstract} 
Studying superconductivity in Dirac semimetals is an important step in understanding quantum matter with topologically non-trivial order parameters. We report on the properties of the superconducting phase in single crystals of the Dirac material \ch{LaCuSb2} prepared by the self-flux method. We find that chemical and hydrostatic pressure drastically suppress the superconducting transition. Furthermore, due to large Fermi surface anisotropy, magnetization and muon spin relaxation measurements reveal Type-II superconductivity for applied magnetic fields along the $a$-axis, and Type-I superconductivity for fields along the $c$-axis. Specific heat confirms the bulk nature of the transition, and its deviation from single-gap $s$-wave BCS theory suggests multigap superconductivity. Our tight-binding model points to an anisotropic gap function arising from the spin-orbital texture near the Dirac nodes, providing an explanation for the appearance of an anomaly in specific heat well below $T_c$. Given the existence of superconductivity in a material harboring Dirac fermions, \ch{LaCuSb2} proves an interesting material candidate in the search for topological superconductivity. 
\end{abstract}

\beginbodyfigures


\section{Introduction} 
In topological semimetals, the electronic band structure features relativistic linearly-dispersive band crossings. This degeneracy gives rise to quasiparticles in condensed matter systems known as Dirac or Weyl fermions, analogous to those found in quantum field theory. Realizing these in bulk materials is an exciting prospect because they bring about topological protection, physical properties beyond those seen in the semiclassical regime, and novel phases of matter. For example, Dirac semimetals are often found to have large linear transverse magnetoresistance, and negative longitudinal magnetoresistance due to the chiral anomaly \cite{Jia, Liang}. Furthermore, new phases of matter can be realized when these topological fermions influence other electronic or magnetic properties, and vice-versa. Exotic topological phases like a monopole superconductor are believed to arise in superconducting magnetic Weyl semimetals with inversion symmetry, whose gap functions are given by monopole harmonics \cite{YiLi}. On the way to realizing such monopole superconductors, it is natural to study centrosymmetric, non-magnetic superconductors that harbor Dirac fermions to understand the physical phenomena that emerge. One family of interest is the set of square-net materials that contain pnictogens like Sb or Bi, where Dirac fermions arise due to nonsymmorphic symmetry \cite{GeunsikLee, Young2015}. Chamorro \textit{et al.}\cite{Chamorro} found a large linear magnetoresistance and small effective masses in the square-net material \ch{LaCuSb2} that were attributed to Dirac fermions. Earlier reports of superconductivity \cite{Gamayunova, Muro} in this material makes it an ideal system to study the interplay between Dirac fermions and superconductivity, however superconductivity was not detected by Chamorro \textit{et al}. 

Here we resolve this contradiction, demonstrating that there is an extreme sensitivity of the physical properties to copper stoichiometry in \ch{LaCuSb2}. We have grown single crystals of \ch{LaCuSb2} with varying copper content to study the effect of disorder on the superconducting state. In optimized samples with the highest superconducting $T_c$ and closest to ideal stoichiometry, we used specific heat, susceptibility, magnetization, and muon spin rotation and relaxation ($\mu$SR) to characterize the superconductivity and its critical behavior. We utilize a combination of simple free-electron models, Bardeen-Cooper-Schrieffer (BCS) models, and tight-binding models to understand the fragile and anisotropic superconductivity in \ch{LaCuSb2}. 

\section{Results}
\begin{figure}
    \centering 
    \includegraphics[width=88mm]{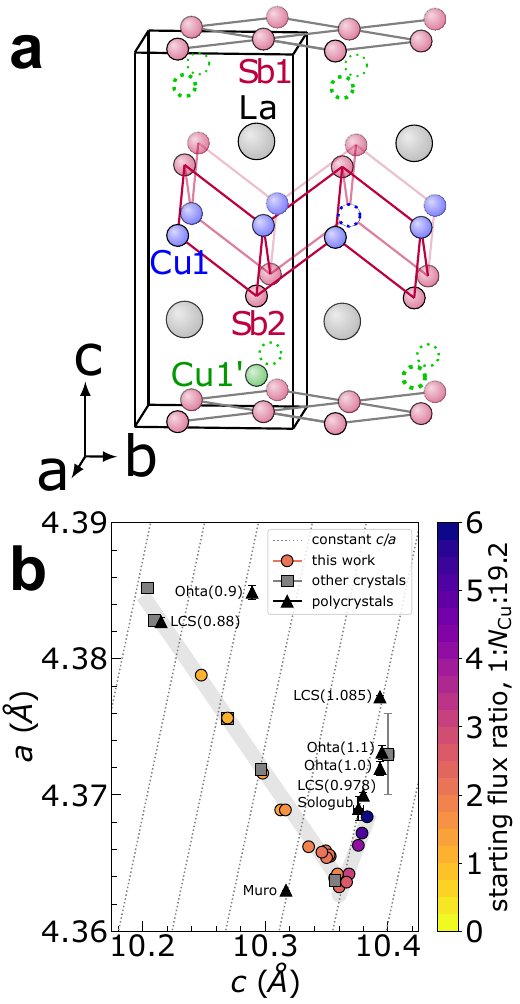} 
    \caption{ \bodyfigurelabel{fig:fig1} Structure and stoichiometry of \ch{LaCuSb2}. (a) Crystal structure of \ch{LaCuSb2}, showing the layered structure and square-net layers of Sb1. The sites Cu1 suffer from partial vacancy, and Cu1' suffer from partial occupancy. (b) Lattice constants $a,c$ determined from powder X-ray diffraction, for various singles crystals and polycrystalline samples from this work and from other works. Lines of constant $c/a$ are used to compare samples in the literature. `LCS($X$)' refer to polycrystal samples from this work, and their refined Cu occupancy $X$ from powder XRD. Data from Muro \textit{et al}.\cite{Muro}, Sologub \textit{et al.}\cite{Sologub}, and Ohta \cite{Ohta} are indicated. Solid grey lines are derived from fitting the data to linear trendlines.} 
\end{figure}

\begin{figure}
    \centering
    \includegraphics[width=88mm]{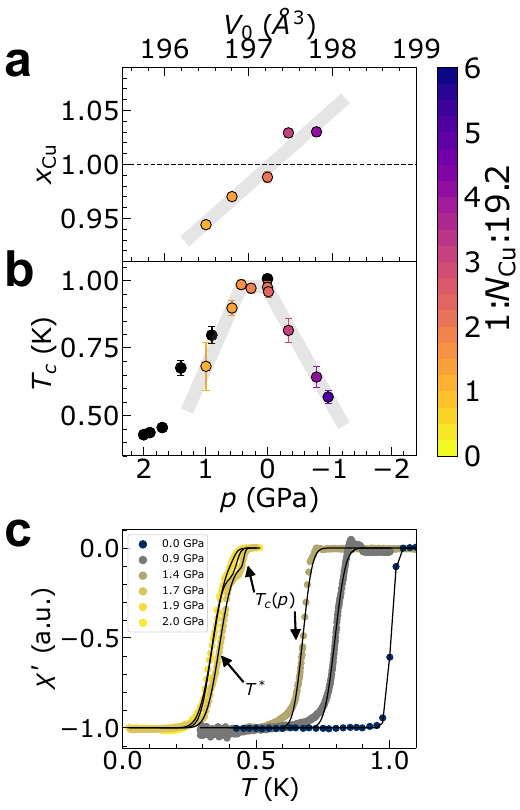} 
    \caption{ \bodyfigurelabel{fig:fig2} Dependence of physical properties on chemical and hydrostatic pressure in \ch{LaCuSb2}. (a) Refined copper occupancy $x_\text{Cu}$ derived from single-crystal XRD, versus room-temperature unit cell volume $V_0=ca^2$ derived from powder XRD. (b) Superconducting dome derived from magnetization measurements, showing the measured midpoint $T_c$ versus $V_0$ at ambient pressure, along with measured midpoint $T_c$ versus hydrostatic pressure $p$ for the optimized sample (black), the latter of which has refined occupancy $x_\mathrm{Cu}=0.988(5)$. Solid grey lines are fits to the data with linear trendlines. (c) Real part of the AC susceptibility for \ch{LaCuSb2} under hydrostatic pressure. The transition $T_c$ is seen to decrease and there appears to be a double-peak feature near $T_c(p)$ and $T^*\approx 0.35$ K at higher pressures. Solid lines are fits to the data using Extended Eq.~\ref{eq:hydrostatic_fit}.} 
\end{figure}

\subsection{Structure} 
\ch{LaCuSb2} crystallizes in the centrosymmetric, nonsymmorphic tetragonal space group $P4/nmm$ (129). It can support both the Cu-deficient ZrCuSiAs structure \cite{Sologub,Ohta}, and Cu-excess defect-\ch{CaBe2Ge2} structure, where the latter features an interstitial Cu site \cite{Ohta,Yang}, as shown in Fig.~\ref{fig:fig1}a. We used $N_\mathrm{Cu}$, related to the flux ratio used for crystal growth, as a parameter to tune stoichiometry in \ch{LaCuSb2} (see Methods). We used powder X-ray diffraction (XRD) to determine the lattice constants of the grown crystals. Fig.~\ref{fig:fig1}b shows the space of lattice parameters $a$ versus $c$ (in \AA) for various crystals at room temperature. There is a clear nearly-linear trend in the $a$ versus $c$ data below and above $N_\mathrm{Cu}\approx 2.25$, where the trend changes from expanding $c$, contracting $a$ to both expanding $a$ and $c$. This is consistent with data from Ohta \cite{Ohta}, whereby the change in the lattice trends occurs at the change between Cu-deficient samples and Cu-excess samples. In the same plot, we have included data from crystals grown with differing starting flux ratios than that reported in this paper. These fall on roughly the same trendlines, affirming that the single parameter $N_\mathrm{Cu}$ can be used to tune a wide range of stoichiometry. We used single-crystal XRD to provide a quantitative estimate of the ratio of the elements, and in particular the copper concentration $x_\mathrm{Cu}$, in various samples, as shown in Fig.~\ref{fig:fig2}a.

\subsection{Superconducting Dome} 
To demonstrate the effect of off-stoichiometry on the superconductivity, we measured DC magnetization on multiple samples grown with different flux ratios. We plotted $T_c$ against parameters characterizing the sample that, over the relevant range of stoichiometry, can be expected to vary if not in proportion then monotonically with the chemical potential. Ours is a self-doping process with Cu1 vacancies in the case of the ZrCuSiAs structure, or occupied Cu1 with interstitial Cu1' ions in the case of the defect-\ch{CaBe2Ge2} structure. Both the unit cell volume $V_0$ and the refined copper occupancy $x_\mathrm{Cu}$ are suitable parameters as a function of which to trace out the superconducting dome. The results are shown in Fig.~\ref{fig:fig2}a and Fig.~\ref{fig:fig2}b, while the supporting susceptibility data are shown in Extended Fig.~\ref{fig:susceptibility}a. We indeed see a systematic change in the superconducting transition temperature as a function of the unit cell and the Cu-occupancy. There is a nearly-linear increase of $T_c$ at small unit cell volumes (Cu-deficient samples), a saturation near nominal stoichiometry until $N_{\text{Cu}} \approx 2$, and a nearly-linear decrease with larger unit cell volumes (Cu-excess samples). 

\subsection{Hydrostatic Pressure} 
Given the large variation in the superconducting transition with chemical composition, we studied the effects of hydrostatic pressure on the superconducting state. As shown in Fig.~\ref{fig:fig2}c, $T_c$ decreases with increasing pressures though full diamagnetic screening is achieved up to the largest pressure of 2 GPa accessed. The reduction in $T_c$ with pressure amounts to $\dv*{T_c}{p} = -0.31(3)$ K/GPa. The suppression of $T_c$ with doping also occurs as the volume decreases at a rate $\dv*{T_c}{v}|_\mathrm{chem} = 80(20)$ K, where $dv=(V-V_0)/V_0$ relative to the optimized $N_\mathrm{Cu}=2$. If the effect of doping were solely associated with chemical pressure, then the corresponding bulk modulus $K=-\dv*{T_c}{v}|_\mathrm{chem}(\dv*{T_c}{p})^{-1}=270(80)$~GPa. With this, the extracted $T_c$ and pressure can be directly compared to the chemical doping, as seen in Fig.~\ref{fig:fig2}b. Note the large effective bulk modulus may be taken as an indication that the effects of copper doping go beyond chemical pressure. Interestingly, for pressures between 1.7 GPa and 2 GPa, there is an anomaly within the transition to diamagnetism near $T^* \approx 0.35$ K. The ambient-pressure specific heat capacity displays a potentially related second lower temperature anomaly near this temperature. 

\subsection{Specific Heat} 
\begin{figure}
    \centering
    \includegraphics[width=\textwidth]{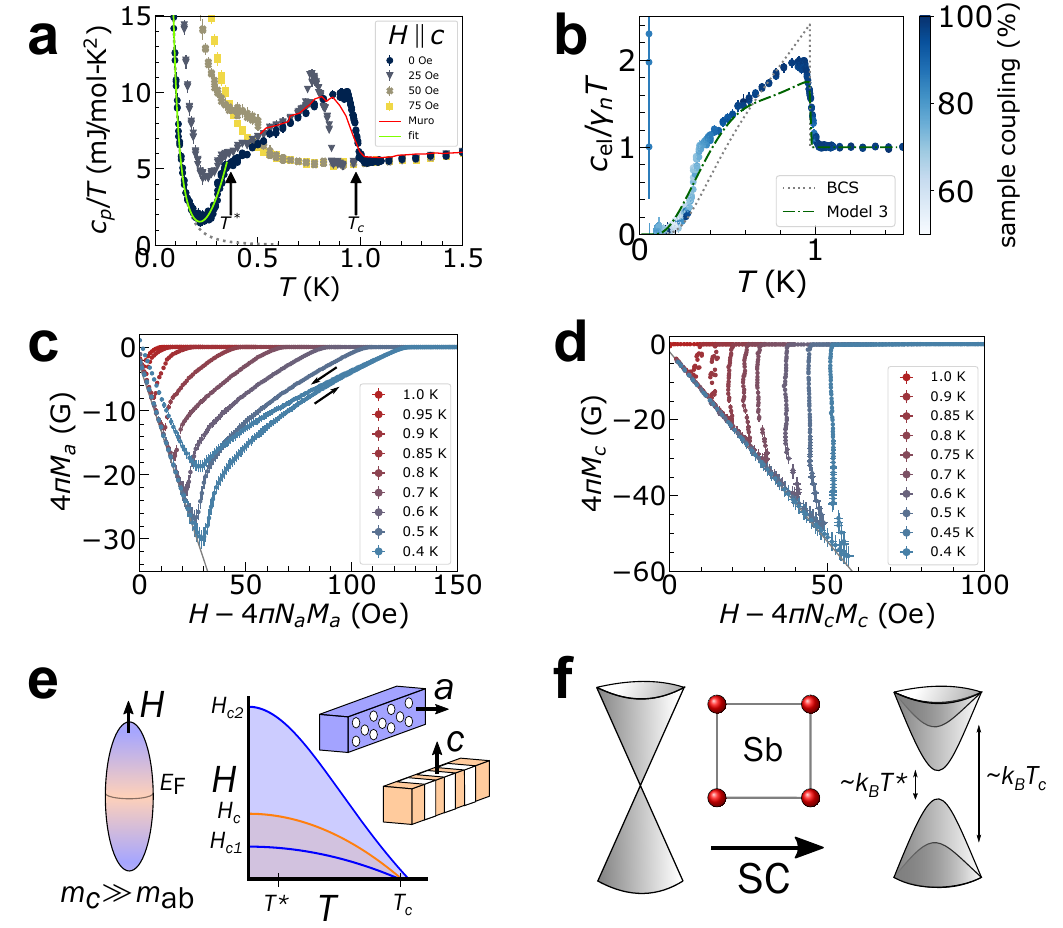}  
    \caption{\bodyfigurelabel{fig:fig3} Thermodynamics of the superconducting state of \ch{LaCuSb2}. (a) Total specific heat $c_p/T$ in applied magnetic field parallel to the c-axis. The superconducting transition temperature $T_c$ and low-temperature anomaly $T^*$ are indicated by the arrows. The fit to the low-temperature zero-field data with Extended Eq.~\ref{eq:SH_fit} is shown in green, and the contribution from the nuclear Schottky is shown as a dotted grey line. The data from Muro \textit{et al.} \cite{Muro} on polycrystal samples are included for comparison. (b) Comparison of electronic specific heat $c_\text{el}/\gamma_n T$ with BCS theory, and the self-consistent two-band Eilenberger model (``Model 3", as discussed in SI, \ref{ssec:SI_GapFit}). (c,d) Magnetization versus internal field, for applied fields along the $a$-axis and $c$-axis, respectively. The dotted grey lines represent the theoretical slope $4\pi \chi_v = -1$, in excellent agreement with the data. (e) The Fermi surface anisotropy and small in-plane effective masses yield an anisotropic field response to the intermediate (Type-I, orange) or vortex (Type-II, blue) state, with normal regions in white. (f) From tight-binding analysis, the spin-orbital texture and superconducting phase transition results in an anisotropic gap that explains the low-temperature anomaly near $T^*$.} 
\end{figure}

To determine whether superconductivity is a bulk effect in \ch{LaCuSb2} and glean information about the superconducting gap function, we turn to specific heat capacity measurements. For a stoichiometric sample in zero field, $C_p/T$ features a sharp jump with midpoint $T_c = 0.98(2)$ K (Fig.~\ref{fig:fig3}a). Apart from a higher transition temperature, $C_p(T)$ is similar to that of polycrystalline samples reported by Muro \textit{et al}. \cite{Muro}. The single crystalline sample, however, enables measurements with a well-defined field orientation. Our measurements of $C_p(T)$ for fields applied along the $c$-axis are in Fig.~\ref{fig:fig3}a. For $H = 75$~Oe superconductivity is fully suppressed, while for $H=25$~Oe the specific heat jump $\Delta C_p$ at $T_c=0.81(4)$ K is actually enhanced over the zero field data. In a Type-I superconductor the superconducting transition becomes first order in an applied field meaning there is latent heat transfer at $T_c$ \cite{Klemm}. While no latent heat was detected for \ch{LaCuSb2} the enhanced value of $\Delta C_p(T_c)/C_p(T_c)= 1.16(3)$ in 25 Oe compared with $\Delta C_p(T_c)/C_p(T_c)= 0.94(4)$ in 0 Oe might be a result of a first-order transition smeared by the inhomogeneous internal field distribution resulting from demagnetization effect in the irregularly shaped sample. 

In contrast to the high-temperature specific heat capacity, which is dominated by phonons and electrons (quantified by the Debye factor $\beta_3$ and Sommerfeld constant $\gamma_n$, respectively), $C_p(T)$ at low $T$ is dominated by a nuclear Schottky anomaly (see SI~\ref{SISpecificHeat} for more details on the corresponding modeling). Subtracting the nuclear Schottky and phonon contributions from the measured specific heat allows us to estimate the zero-field electronic specific heat, $C_\text{el}(T) = C_p(T) - C_N(T) - \beta_3 T^3$, shown in Fig.~\ref{fig:fig3}b. From these data, we conclude that superconductivity in \ch{LaCuSb2} is of bulk origin and not a secondary phase or surface effect. The electronic specific heat capacity also may be directly compared with predictions from BCS theory. For instance, the size of the specific heat jump in zero field $\Delta c/\gamma_n T_c = 0.94(4)$ is less than that expected from BCS theory of $\Delta c/\gamma_n T_c = 1.43$. This suggests a gap $\Delta(0)/\gamma_n T_c$ differing from conventional BCS theory, or possibly multiband superconductivity. Furthermore, the electronic specific heat features an abrupt drop at low temperatures near $T^* \approx 0.35$ K, leading to an exponentially-activated decline. This is a similar temperature scale to the kink seen in the high-pressure susceptibility data in Fig.~\ref{fig:fig2}c. As we will show, our tight-binding model points to the possibility of an anisotropic gap function that causes the feature at $T^*$. To highlight the main features of the data, we explore several possible models in the SI \ref{ssec:SI_GapFit}, including the Eilenberger self-consistent two band model \cite{Prozorov2}, shown in Fig.~\ref{fig:fig3}b.

\subsection{Quantum oscillations}
To quantify changes in the Fermi surface, we studied Shubnikov-de Haas quantum oscillations in the Hall resistivity for a stoichiometric sample. The large residual-resistivity ratio of 16 indicates high sample quality (see SI, \ref{SI_Transport}). The magnetoresistance $\rho(H)/\rho(0)$ is nearly linear in fields up to 16 T (Fig.~\ref{fig:QO}a), and is largest at the lowest temperatures, similar to previous results found for this Dirac material \cite{Chamorro}. Next, the Hall data (Fig.~\ref{fig:QO}b) show nearly linear behavior at low fields less than 7T, but reveals non-linearity and quantum oscillations at high fields up to 16 T. We extracted the oscillatory behavior as a function of field and temperature from the Hall data (Fig.~\ref{fig:QO}c). Using the amplitude of the 48 T frequency and the Lifshitz-Kosevich formula \cite{Shoenberg}, we could extract the effective masses of the charge carriers (Fig.~\ref{fig:QO}d). This is dominated by small effective masses that mainly arise from the small electron pockets near the $X$ point. We find $m_a^* = 0.058 \; m_e$ from the 48 T oscillation, which is within error of the masses found in samples from Chamorro \textit{et al.} \cite{Chamorro} and Akiba \textit{et al.} \cite{Akiba2023}. It is worth noting the 48 T frequency found here is relatively unchanged compared to the above works, suggesting self-doping has negligible effects on the Fermi surface geometry. 

\begin{figure}
    \centering
    \includegraphics[width=\linewidth]{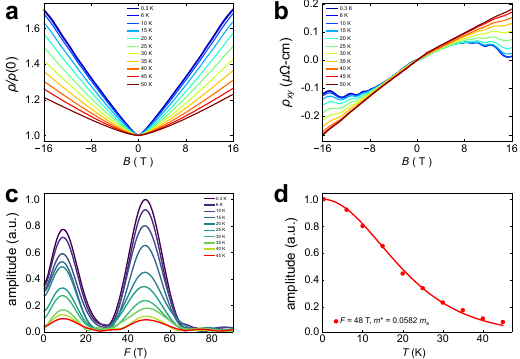}
    \caption{ \bodyfigurelabel{fig:QO} Quantum oscillations in the optimal $N_\mathrm{Cu}=2$ samples. (a) Magnetoresistance $\rho(H)/\rho(0)$ at various temperatures from 0.3 K to 50 K and for applied magnetic fields up to 16 T. (b) Hall effect for fields up to 16 T, showing nonlinearity and quantum oscillations in high fields. (c) Amplitude of the quantum oscillations as a function of frequency, for various temperatures. Note the peak near 10 T may be an artifact of the background removal. (d) Amplitude of the 48 T peak as a function of temperature, fit with the Lifshitz-Kosevich formula to extract the effective cyclotron mass.} 
\end{figure}

\subsection{Anisotropic Magnetization} 
We turn to magnetization measurements on the optimized sample to study the critical fields in \ch{LaCuSb2}. Demagnetization corrections were taken into account to allow us to report the internal magnetic field $H_{\mathrm{int}}$ (see SI, \ref{Mag_and_Susc}). The $4\pi M$ versus $H_{\mathrm{int}}$ data can distinguish between Type-I and Type-II superconductivity. Fig.~\ref{fig:fig3}c,d shows the demagnetization-corrected magnetization data for internal fields along the $a$- and $c$-axis, respectively. Note the slopes $\chi = dM/dH_\mathrm{int}$ are consistent with the expected value for bulk superconducting susceptibility $4\pi \chi = -1$ in both cases. With applied fields along the $a$-axis, the magnetization is linear at small fields indicative of a Meissner state, but then the diamagnetic magnetization abruptly decreases for $H>H_{c1}(T)$. This suggests Type-II superconductivity with a small critical field $H_{c2}(0) = 172(6)$ Oe, as deduced from extrapolating the data to zero temperature (see SI, \ref{Mag_and_Susc}). For applied fields along the $c$-axis, the magnetization remains linear for an extended field range until it sharply drops to zero magnetization near a critical field, rather than gradually decaying due to the appearance of flux lines as in the previous case, indicative of Type-I superconductivity. The response of \ch{LaCuSb2} to magnetic fields is highly anisotropic, as shown schematically in Fig.~\ref{fig:fig3}e.

\begin{figure}
    \centering
    \includegraphics[width=\textwidth]{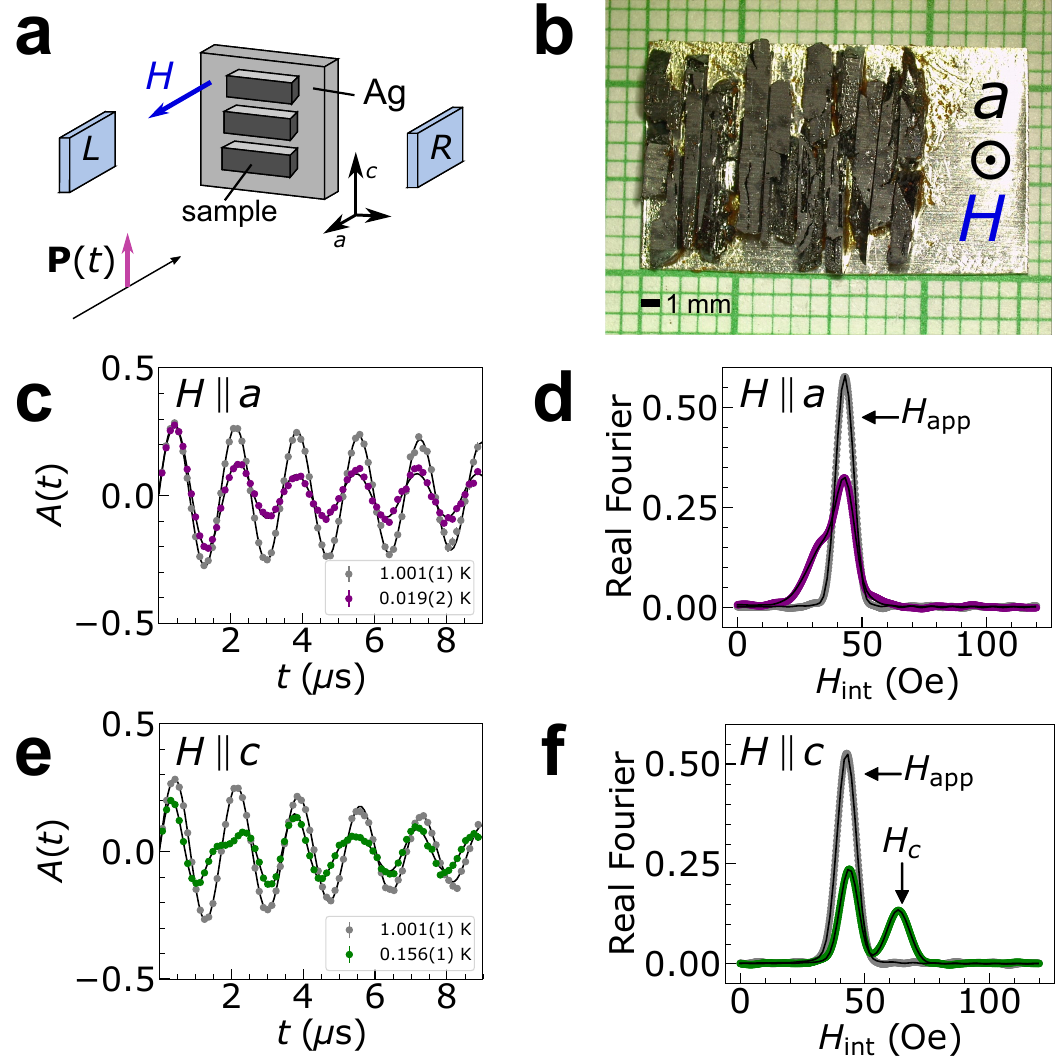} 
    \caption{ \bodyfigurelabel{fig:fig4} Transverse-field $\mu$SR investigation on \ch{LaCuSb2}. (a) Experimental set-up for the $\mu$SR experiment. The muon spin is flipped to be perpendicular to the incident beam, and the magnetic field $H_\mathrm{app}$ is applied horizontally. The left/right detectors were used for the experiment. (b) Image of the co-mounted samples mounted with the $a$-axis out of the page (in the same direction as $H_\mathrm{app}$) and with the $c$-axis horizontal. (c) Muon asymmetry and (d) real Fourier amplitude in the Type-II superconducting state for $H_\mathrm{app}\approx 40$ Oe applied along the $a$-axis. There are peaks at the applied field, and smaller internal fields experienced by the muons in the vortex state. Solid lines are fits to Eq.(\ref{eq:TF_muSR_a-axis}).  (e) Muon asymmetry and (f) real Fourier amplitude in the Type-I superconducting state for $H_\mathrm{app}\approx 40$ Oe applied along the $c$-axis. There are peaks at the applied field, and a larger internal field experienced by the muons in normal regions. Solid are fits to Eq.(\ref{eq:TF_muSR_c-axis}).} 
\end{figure}

\subsection{Transverse-field $\mu$SR} 
So far, our discussions have involved measurements that probe bulk sample-averaged thermodynamic properties. The use of a local probe like $\mu$SR complement these results and confirm the anisotropic superconducting nature of \ch{LaCuSb2}. We used transverse-field (TF) $\mu$SR to study the response of the superconducting state to applied magnetic fields. The geometry of the experiment is shown in Fig~\ref{fig:fig4}a, with the co-mounted samples shown in Fig~\ref{fig:fig4}b. When the samples were co-aligned with magnetic field parallel to the $a$-axis, the real Fourier transform amplitude (Fig.~\ref{fig:fig4}d) of the asymmetry data (Fig.~\ref{fig:fig4}c) showed a broad peak centered at fields lower than the applied field. The broad distribution of internal fields sampled by the muon ensemble indicates the formation of a vortex lattice. We fit the asymmetry data to 
 \begin{equation} \label{eq:TF_muSR_a-axis}
    A_a(t) = A_0 [F e^{-\sigma^2 t^2/2} \cos(\omega t +\phi) + (1-F) e^{-\lambda_\mathrm{bg} t} \cos(\omega_\mathrm{bg} t +\phi)]. 
 \end{equation} 
Here $F$ is the fraction of muons that stop in \ch{LaCuSb2} with the remainder stopping in the Ag sample holder.  $\omega = \gamma_\mu H_\mathrm{int}$ is the muon precession frequency in the average internal field; $\omega_\mathrm{bg} = \gamma_\mu H_\mathrm{app}$ is the corresponding frequency for muons that stop in the Ag sample holder. $\lambda_{bg}$ describes the exponential relaxation rate for muons that stop in Ag. $\sigma$ characterizes the width of the internal field distribution in the sample; $\gamma_\mu = 2\pi \cdot 135.5$ MHz/T is the gyromagnetic ratio of the muon; and $\phi$ is the phase angle of the initial muon polarization. For the purpose of our analysis, the gaussian relaxation term in Equation~\ref{eq:TF_muSR_a-axis} provides an adequate approximation to the Fourier transform of the internal field distribution $n(H)$ for the vortex lattice, which peaks at fields $H<H_\mathrm{app}$ \cite{Sonier}. With certain assumptions (see SI, \ref{ssec:TF_muSR}), we used the temperature dependence of the relaxation rate to obtain the superfluid density $\rho(T) = \lambda^2(0)/\lambda^2(T)$. This may be compared to theoretical models to extract information about the gap function, and possible multiband effects (see Extended Fig.~\ref{fig:LCS_Models} for fitting models and their discussion). Most notably, we find the low temperature limit is consistent with conventional $s$-wave behavior and no thermal anomaly is observed  in $\rho(T)$ near $T^*$.

We then measured the same crystals with the magnetic field oriented parallel to the $c$-axis. In stark contrast to the previous case, the real Fourier transform amplitude (Fig.~\ref{fig:fig4}f) of the asymmetry data (Fig.~\ref{fig:fig4}e) featured a broad peak at internal fields higher than the applied field. To understand this, we note that in the Meissner state, most muons will enter the superconductor where $B=0$, while muons implanted in the Ag sample holder precess in the applied field. However, due to demagnetization effects, some regions of the superconducting sample sustain an internal field that is greater than the critical field $H_{c}$. In Type-I superconductors, such a region must become normal, as there is no other state (such as a vortex state) to support superconductivity. A Type-I superconductor with demagnetization factor $N$ in an applied field $H_\mathrm{app} > (1-N)H_{c}$ is in the intermediate state, characterized by laminar structures of superconducting and normal regions, the latter of which maintains a constant internal field equal to the critical field $H_{c}$ \cite{Beare,Kozhevnikov}. In our $\mu$SR experiment, $N\approx 0.86$ for the plate-like ensemble of co-aligned single crystals, and $H_{c} = 59.8(1.0)$ Oe from magnetization data, meaning that in applied magnetic fields above $H_\mathrm{app} \approx 8.4$ Oe, the samples were assuredly in the intermediate state. Therefore, the observed oscillation frequency in the $\mu$SR spectrum corresponds to the critical field $H_{c}$. To model the data and extract $H_{c}(T)$, we fit the asymmetry spectrum using the equation 
 \begin{equation} \label{eq:TF_muSR_c-axis}
    A_c(t) = A_0 [F (1-F_S) e^{-\sigma^2 t^2/2} \cos(\omega t +\phi) + (1-F) e^{-\lambda_\mathrm{bg} t} \cos(\omega_\mathrm{bg} t +\phi)]
 \end{equation}
Here $F$ as before is the fraction of muons stopping in the sample, and $F_S$ is the superconducting volume fraction of that sample. Only the normal regions with volume fraction $F(1-F_S)$ will have a non-zero frequency $\omega$ due to the intermediate state. From the fits to the data, we find a large superconducting volume fraction $F_S = 0.924$ at 20 mK and 10 Oe, indicative of a bulk superconducting response as we expect from the thermodynamic data. We also extracted the critical field $H_{c}(T)$ by tracking the temperature-dependence of the frequency $\omega>\omega_\mathrm{bg}$ from the normal regions in the sample. We found that in this entire temperature range (down to 0.02 K) in 10 Oe and 40 Oe, the FFT spectrum of muon precession always showed a peak at a field higher than the applied field. This provides microscopic evidence that \ch{LaCuSb2} is in a pure Type-I superconducting state when magnetic fields are applied parallel to the $c$-axis. Combining the $a$-axis and $c$-axis oriented magnetic fields allows us to map out the phase diagram, with the vortex behavior (and lack thereof) sketched in Fig.~\ref{fig:fig3}e and demonstrated quantitatively in Extended Fig.~\ref{fig:phase_diagram}.

\subsection{Tight-Binding Model and Superconductivity} 
Towards a microscopic understanding of superconductivity in \ch{LaCuSb2}, we implemented a tight-binding model for the pocket near the $X$-point that describes multiple bands derived from Sb orbitals related by the space-group symmetry \cite{Schoop2016,Young2015}. The derivation is described in detail in the SI~\ref{SITB}. The pair-wise degeneracy of the bands at $X$ is enforced by the nonsymmorphic symmetry of the space group. These bands intertwine across the Brillouin zone, which contributes additional spin-orbital texture near $X$ point. Assuming local isotropic attractions, we find anisotropic superconducting gaps on the Fermi pockets around $X$. We tentatively attribute the anomaly $T^*$ observed in the specific heat measurement to this extreme anisotropy of the pairing gaps. At lowest temperatures, the excitations are exponentially-suppressed due to nodeless superconducting gaps, as in a conventional $s$-wave superconductor. When the temperature increases and reaches the smallest gap size, thermal fluctuations are significantly enhanced, giving rise to the anomalous increase in the specific heat observed around $T^*=0.35$ K. However, the superconducting gaps do not close at $T^*$, so this is a thermal cross-over phenomenon and there is no peak in the specific heat. When the temperature reaches $T_c$, the superconducting gaps close, which gives rise to the sharp discontinuity at $T_c$ as usual for a BCS superconductor. This picture is sketched in Fig.~\ref{fig:fig3}f. 

\section{Discussion} 
Despite the anisotropy we observe experimentally, we gain quantitative insight into the superconducting state by calculating thermodynamic quantities in a one-band, isotropic free-electron model as a first approximation. The relevant thermodynamic and superconducting quantities are tabulated in Table~\ref{tab:SC_quantities}. In particular, we find a coherence length of $\xi\approx 1\mu$m, whereas the mean-free path is $\ell\approx 0.06 \mu$m, suggesting \ch{LaCuSb2} is in the dirty limit\cite{Tinkham} since $\xi \gg \ell$. The relevant expression \cite{Weber} for calculating the Ginzburg-Landau (GL) parameter in the dirty limit is $\kappa = (7.49\times 10^3) \gamma_{nV}^{1/2} \rho_{0}$ (with all quantities in cgs units), where $\rho_0$ is the value of the low-temperature residual resistivity plateau. The out-of-plane GL parameter $\kappa_c$ depends on currents generated in the $ab$ plane, so using $\rho=\rho_{0a}$ for in-plane resistivity, we find $\kappa_c = 0.398(7)$. That $\kappa_c<1/\sqrt{2}$ is consistent with Type-I superconductivity for $H\parallel c$. Extrapolating our magnetization data to $T=0$, we have $H_c = 59.8(1.0)$ Oe and $H_{c2} = 172(6)$ Oe and thereby estimate\cite{Tinkham} $\kappa_a = H_{c2}/\sqrt{2} H_{c} = 2.03(8)$. Next, we use conventional BCS theory to estimate the thermodynamic critical field $H_{c}$. This field is related to the gap function and density of states \cite{Itoh}, and in cgs units is given by $H_c(0) = 1.764\sqrt{6/\pi} \cdot \gamma_{nV}^{1/2} T_c$. From this we estimate that $H_c(0) \approx 68(1)$ Oe, in remarkable agreement with the measured phase diagram (see Extended Fig.~\ref{fig:phase_diagram}). We attribute the small critical field to the low $T_c$ and the small density of states in this Dirac material. 

Although Type-I superconductivity is mainly found in elemental superconductors and several binary compounds (e.g. noncentrosymmetric \ch{BeAu} \cite{Beare} and Dirac semimetal candidate \ch{PdTe2} \cite{Leng2019}), it has also been observed in ternary compounds including \ch{LaRhSi3} \cite{Anand} and \ch{LiPd2Ge} \cite{Gornicka}, among others. However, anisotropic Type-I and Type-II superconductivity is not as widely reported. In the theory of conventional anisotropic superconductors \cite{Klemm,Kogan2002}, the anisotropy factor $\gamma = \kappa_a/\kappa_c$ is related to the effective masses as $\gamma^2 = m_c/m_a$. The anisotropy of the GL parameter is directly related to the anisotropy of the Fermi surface, as schematized in Fig.~\ref{fig:fig3}e. Furthermore, it is at the same time possible to satisfy that $\kappa_c<1/\sqrt{2}$ and $\kappa_a > 1/\sqrt{2}$ when $\gamma\gg 1$, implying a superconductor whose type depends on the applied magnetic field direction \cite{Klemm}. In particular, the direction of the field relative to the crystallographic axes will either favor or disfavor vortices in the superconducting order parameter, based on the free energy in distinct crystallographic directions. While rare, this behavior has been observed in \ch{C8K} \cite{Koike} and TaN \cite{Weber,Sporna}, where the angular dependence of the critical field was explicitly studied. However, such an anisotropy seems not to be well-studied in ternary systems. Considering the previous estimates of $\kappa_a$ and $\kappa_c$, the large anisotropy parameter $\gamma^2 \approx 26(2)$ is a consequence of the small in-plane effective masses of this Dirac material. 

From specific heat, we have pointed to the possibility of multigap behavior or an anisotropic gap, supported by the findings of our tight-binding model. However, determining the temperature-dependence of the gaps using our Hamiltonian requires fitting many free parameters. In Extended Fig.~\ref{fig:LCS_Models}, we highlight several toy models that capture the overall features in specific heat and superfluid density, but do not capture all features. One possibility to measure this anisotropic gap is using quasi-particle interference with Scanning Tunneling Microscopy (STM). The quasiparticle interference signal should be significantly enhanced near the gap edge, which may be directly compared to the anomaly in the specific heat. We find it quite compelling to obtain a more complete microscopic understanding of the superconducting state of \ch{LaCuSb2} using a combination of theory and probes like STM. Overall, despite the fragility of the superconducting state in \ch{LaCuSb2} to stoichiometry, pressure, and magnetic fields, it is worth exploring related compounds, perhaps ones that contain magnetic moments, to continue the search for exotic topological phases like the monopole superconductor. 

\section{Author Contributions} 
Sample synthesis and data analysis on the superconducting state was performed by C.J.L. under the supervision of C.L.B. and T.M.M. The $\mu$SR experiments were carried out by C.J.L., J.G., M.P., S.S., and P.O. under the supervision of C.L.B and G.M.L. The tight-binding analysis was performed by J.Z. under the supervision of Y.L. The DFT calculations for the band structure were performed by S.G. The heat capacity measurements were performed by C.J.L. and T.H. The He3 susceptibility and magnetization measurements were performed by C.J.L., T.B. and J.R.C. Quantum oscillations were measured and analyzed by G.G. under the supervision of B.J.R. Single-crystal X-ray diffraction measurements and refinements were performed by M.A.S. TEM measurements were performed by K.L. The hydrostatic pressure measurements were performed by K.M. and D.K.B. under the supervision of Y.U. and S.N. All authors contributed with comments and edits. 

\begin{methods}


\textbf{Synthesis}: Single crystals of \ch{LaCuSb2} were grown by the self-flux method. Cut pieces from lanthanum ingot (Ames Laboratory, 99.99\%), pieces of copper (Alfa Aesar, 99.999\%), and antimony shot (Strem Chemicals, 99.9999\%) were weighed with a total mass of about 4 g in various molar ratios and placed in an alumina crucible. An inverted catch crucible with an alumina strainer was placed atop the first crucible, and both were sealed in a quartz tube under partial pressure of argon gas. The ampules were heated to 1070$^\circ$ C at a rate of 100$^\circ$ C/h and held for 12 h, then cooled to 670$^\circ$ C at a rate of 4$^\circ$ C/h before centrifuging. Inspired by previous literature on isostructural \ch{LaAgSb2}, which was reported to produce stoichiometric samples with starting ratio $0.045:0.091:0.864$ (or roughly, $1:2:19.2$) \cite{Myers, Masubuchi}, we used varying compositions $1:N_\text{Cu}:19.2$ with $1 \leq N_\text{Cu} \leq 6$ as a parameter to tune the chemical potential of the system and grow crystals with different Cu content. 

Polycrystalline samples were also prepared using a simple reaction of the elements in a quartz tube. Powders with ratios $1:\delta:2$ with $\delta = 0.8, 1.0, 1.2$ were synthesized by first melting about 1.0 g of the elements in a quartz tube, using a step furnace at 600$^\circ$ C for 24 hours. This polycrystal was then ground and reheated to $800^\circ$C for roughly 12 hours. Further heating of the powder at high temperatures, or low temperatures for extended periods of time, was disadvantageous due to the decomposition of the structure, as evidenced by larger relative percentages of secondary phases \ch{Sb} and \ch{Cu2Sb}. We compared the lattice constants (in particular, the $c/a$ ratio) of the polycrystalline samples to those of our single crystals, and to those from the literature. 

\textbf{X-ray diffraction}: Powder XRD data were collected at room temperature using a laboratory Bruker D8 Venture Focus diffractometer with LynxEye detector in the range from 10-80 degrees. Refinements on the powder XRD data were performed using Topas 5.0 (Bruker). The occupancy of La and Sb ions were constrained to 100\%, while the Cu occupancy was refined freely. We also refined the strain parameter and preferred orientation, due to the layered nature of the crystals. 

Single-crystal XRD data were acquired at 110(2) K using a SuperNova diffractometer (equipped with Atlas detector) with Mo K$\alpha$ radiation ($\lambda = 0.71073$ \AA) under the program CrysAlisPro (Version CrysAlisPro 1.171.39.29c, Rigaku OD, 2017). The same program was used to refine the cell dimensions and for data reduction. The structure was solved with the program SHELXS-2018/3 and was refined on F2 with SHELXL-2018/3 \cite{Sheldrick}. Analytical numeric absorption correction using a multifaceted crystal model was applied using CrysAlisPro. The temperature of the data collection was controlled using the system Cryojet (manufactured by Oxford Instruments). For all samples in the ZrCuSiAs structure type, the occupancy factor for Cu1 was refined freely. For samples in the defect-\ch{CaBe2Ge2} structure type, the additional Cu site (denoted Cu1') was necessary to obtain good fits to the data. The occupancy factor for both Cu1 and Cu1' were refined freely, and the reported occupancy is $x_\mathrm{Cu} = x\mathrm{[Cu1]}+x\mathrm{[Cu1']}$. Crystallographic data tables are included in the Supplementary Information. 

\textbf{Specific heat}: Heat capacity measurements were performed in a Quantum Design Physical Properties Measurement System (PPMS) with the dilution refrigerator option. The magnetic field was degaussed above 3.8 K to minimize effects of persistent fields. Measurements were performed on a single crystalline sample of LaCuSb$_2$ of 10.7(1) mg mass oriented such that the applied magnetic field was along the nominal $c$-axis, with reported fields depicted in Fig.~\ref{fig:fig3} between 0 Oe and 75 Oe. All measurements were performed in a fixed magnetic field and measured upon cooling, with a minimum temperature between $T=0.05-0.1$ K and a maximum temperature of $T=3.8$ K. 

\textbf{Magnetization}: Magnetization measurements were performed in a Quantum Design (QD) Magnetic Properties Measurement System (MPMS) with QD iHelium3 He3-insert. To extract the superconducting dome, we measured samples with nominal applied fields of 2 Oe along the $a$-axis, after degaussing at temperatures above $T_c$. The samples were cut with a large aspect ratio along the $a$-axis such that the demagnetization corrections were small. Measurements on the optimal sample were performed taking into account the non-zero demagnetization factors. We performed isothermal measurements in applied field by first degaussing at high temperature above $T_c$. We then cooled in zero field to the appropriate temperature, and measured from zero to fields well past the point that magnetization vanished to study $4\pi M$ versus $H_\mathrm{int}$. 

\textbf{Electrical transport}: Shubnikov-de Haas oscillations were measured in a single crystal cut from the same crystal used for other thermodynamic measurements, at temperatures down to 0.3 K and fields up to 16 T. Resistivity and Hall effect measurements above 2K were performed on a PPMS using the AC Transport (ACT) option. For all resistivity measurements we prepared a polished bar-shaped crystal with the current applied along the $a$-axis, using a four-probe configuration consisting of platinum wires and Epo-Tek silver epoxy. For Hall effect measurements we used the five-probe method which allowed the Hall signal to be tuned to zero in zero applied field. Low-temperature resistivity measurements were performed in a PPMS using a Lake Shore Model 372 AC Resistance Bridge. We applied an AC current of 316 $\mu$A at a frequency of 13.7 Hz. The field was applied along the $c$-axis after degaussing the magnet at 1.1 K (above the transition) to reduce the effect of persistent magnetic fields. Measurements were taken upon cooling in various applied fields. 

\textbf{Hydrostatic pressure}: Measurements under hydrostatic pressure were performed using a Bluefors dilution refrigerator down to 0.05 K and up to 2.0 GPa at the Institute for Solid State Physics (ISSP), The University of Tokyo. To track the variation of the superconducting transition temperature under hydrostatic pressure, the AC magnetic susceptibility of a sample was measured with a mutual induction method at a fixed frequency of 317 Hz with a modulation field of about 1 Oe.  Measurements were performed on the optimized samples $N_\mathrm{Cu} = 2$, cut from the same crystal used for magnetization, specific heat, and resistivity measurements. For applying pressure, a piston-cylinder cell made from nonmagnetic BeCu and NiCrAl alloys was used with Daphne 7373 as the pressure transmitting medium. The pressure was determined from the superconducting transition temperature of Pb. 

\textbf{Muon spin rotation}: Zero-field and transverse-field muon spin rotation and relaxation measurements were performed at the TRIUMF facility in Vancouver, Canada. A spectrometer incorporating a dilution refrigerator was used on the M15 beamline, to allow for measurements down to 20 mK. The setup makes use of a superconducting magnet to allow for magnetic fields up to 5 T, and resistive coils for finer control and field-zeroing. The magnetic field was applied horizontally, parallel to the direction of the muon beam. In the ZF measurements, the muon spin was (anti)parallel to the beam direction, while in TF measurements the muon spin was perpendicular to the field and beam direction. Single crystals grown with the optimal ratio $N_\mathrm{Cu}=2$ and with the greatest thickness (average 1.0 mm) were cut along the $a$-axis. We used multiple co-aligned single crystals, totaling a mass of about 0.98g, to maximize the measured signal and reduce background from muons not implanted in \ch{LaCuSb2}. The crystals were first mounted with the $a$-axis parallel to the applied field for the first measurement, and the same crystals were then individually rotated and remounted with the $c$-axis parallel to the applied field for the second measurement. The samples were mounted on a silver cold finger using a mixture of Apezion N grease and copper-loaded Cry-Con grease to ensure good thermal contact. For the field along the $a$-axis, we used $H_\mathrm{app} = 40$ Oe. For the field along the $c$-axis, we used $H_\mathrm{app} = 40$ Oe to access low critical fields and $H_\mathrm{app} = 10$ Oe to access higher critical fields. All data at constant field were simultaneously refined and fit for various temperatures using the \textit{musrfit} program. \cite{Suter}
\end{methods}

\noindent{\bfseries References}\setlength{\parskip}{12pt}%

\bibliographystyle{naturemag}



\begin{addendum}
\item [Acknowledgments] This work was supported as part of the Institute for Quantum Matter, an Energy Frontier Research Center funded by the U.S. Department of Energy, Office of Science, Office of Basic Energy Sciences, under Award DESC0019331. C.J.L. acknowledges the support of the William Gardner Fellowship. The 3He MPMS was funded by the National Science Foundation, Division of Materials Research, Major Research Instrumentation Program, under Award 1828490. Research at McMaster University was supported by the Natural Sciences and Engineering Research Council (GML). The high pressure measurements were funded by KAKENHI (no. JP19H00648), JST-Mirai Program (no. JPMJMI20A1), JST-CREST (no. JPMJCR18T3 and JPMJCR15Q5). We would like to thank Bassam Hitti and Sarah Dunsiger for their assistance during the $\mu$SR experiment; Lisa Pogue for her preliminary DFT work; and Yishu Wang for her assistance with resistivity measurements. 

 \item[Competing Interests] The authors declare that they have no competing financial interests.
 \item[Correspondence] Correspondence and requests for materials should be addressed to clygour1@jhu.edu 

 \item[Note added] While writing this paper, we became aware of an experimental study on single crystals of \ch{LaCuSb2} \cite{Akiba2023}. The specific heat, resistivity, and Hall effect measurements are consistent with our data. However, no investigation into the magnetic field anisotropy was reported. Furthermore, the presence of a low-temperature feature in thermodynamic measurements allows us to gain insight on the contributions from the Dirac band structure.
\end{addendum}



\processdelayedfloats
\setcounter{mypostfigure}{0}
\setcounter{figure}{0}
\setcounter{page}{1}
\setcounter{section}{0}

\beginedfigures

\newpage

\noindent{\bfseries \LARGE Supplementary Information}\setlength{\parskip}{12pt}%

\section{Crystal growth and effects of stoichiometry}

The flux-grown crystals were plate-like with average surface areas of 1 cm$^2$, often limited by the crucible diameter, and with varying thickness. The least Cu-rich fluxes produced crystals with the most appreciable thickness, compared to the most Cu-rich fluxes having large surfaces areas and decreased thickness. Grown crystals with copper-rich fluxes often came out with layers of solidified \ch{Sb}-\ch{Cu2Sb} flux on the surface, which could be removed with mechanical polishing. The presence of these phases is expected from the ternary phase diagram reported previously \citesec{Gschneidner}. Inclusions, primarily Sb and less-so \ch{Cu2Sb}, were present in all samples, and Sb was used as a reference for the lattice constants for powder XRD. Importantly, \ch{Sb} is not superconducting at ambient pressure \citesec{Wittig} and \ch{Cu2Sb} is only superconducting below 0.085 K \citesec{Andres}, and thus cannot account for the superconducting signals observed near 1 K in our samples. However, the nature of off-stoichiometry may effect the physical properties we observe. For example, removal of entire planes of Cu would in principle shrink the unit cell to an effective \ch{LaSb2} structure, which (in the orthorhombic structure) is known to be superconducting \citesec{Guo, Ruszala}

The fact that $N_\mathrm{Cu}$ can tune a wide range of compositions is likely due to the change of thermodynamic chemical potential and stable phases that can emerge in the \ch{LaCu_xSb2} solid-solution-type structure. If the main role of the Cu off-stoichiometry is interpreted as the introduction of non-magnetic impurities (as opposed to stacking defects, see TEM results below), in the form of vacancies or interstitial sites, then according to Anderson's theorem on dirty s-wave superconductors, such defects should have no effect on $T_c$ \citesec{Anderson}. There may be several ways to reconcile this. One possibility is that the small effects of off-stoichiometry may be affecting the Fermi level and density of states, as suggested in resistivity measurements (see \ref{SI_Transport}), thereby resulting in stronger dependence of $T_c$ as expected under conventional BCS theory \cite{Tinkham}. Another possible explanation is that the gap function itself is anisotropic leading to strong dependence of $T_c$ on the scattering rate, a well-known result from Gor'kov \citesec{Petrovic, Gorkov}. Indeed, an anisotropic gap function is expected from our tight binding model for the pairing gap function. It is possible that one or both effects could explain the drastic suppression of $T_c$ found in the superconducting dome.

\section{Transmission Electron Microscopy (TEM)}
\label{TEM} 

A Cu-deficient sample with $N_\mathrm{Cu} = 1$, and a near-stoichiometric sample with $N_\mathrm{Cu} = 2$, were ground into a fine powder for use in TEM. The samples used were from the same crystals cut for thermodynamic measurements. Microcrystallites nearly oriented along $a$ and $c$ were studied for their fringe patterns to estimate the lattice constants, and to gauge small variations in fringe width, which may give evidence of stacking defects in the unit cell. This high-resolution TEM data precludes variations in the lattice constant within $\pm 0.44$\AA$\;$ in 27 unit cells. We used the GMS-3 software to integrate over regions shown in Extended Fig.~\ref{fig:TEM}, and extract the intensity as a function of position over the length of the rectangles. Near each maxima, we fit the location of the central peak using a simple quadratic fit. This was used to deduce the distribution of distances between successive fringes. The resulting histogram shows spread within one pixel, i.e. a spread of 0.44 \AA. The lack of clear bimodal distributions outside the resolution of the scan indicates there are no obvious stacking disorders that may produce \ch{LaSb2}. Given the off-stoichiometry on the order of 0-5\%, this suggests possibly statistical vacancies of Cu atoms as suggested in \citesec{Klemenz}. Interestingly, the samples with smaller copper flux ratios $N_\mathrm{Cu} \approx 1$ that were cut into the bulk of the sample appeared to have a copper-colored tint on the exposed cut surface after a period of several months. This indicates the copper ions may be mobile in the structure at room temperature, made possible by the presence of vacancies and a path by which ions can travel. 
\begin{figure}
    \centering
    \includegraphics[width=\linewidth]{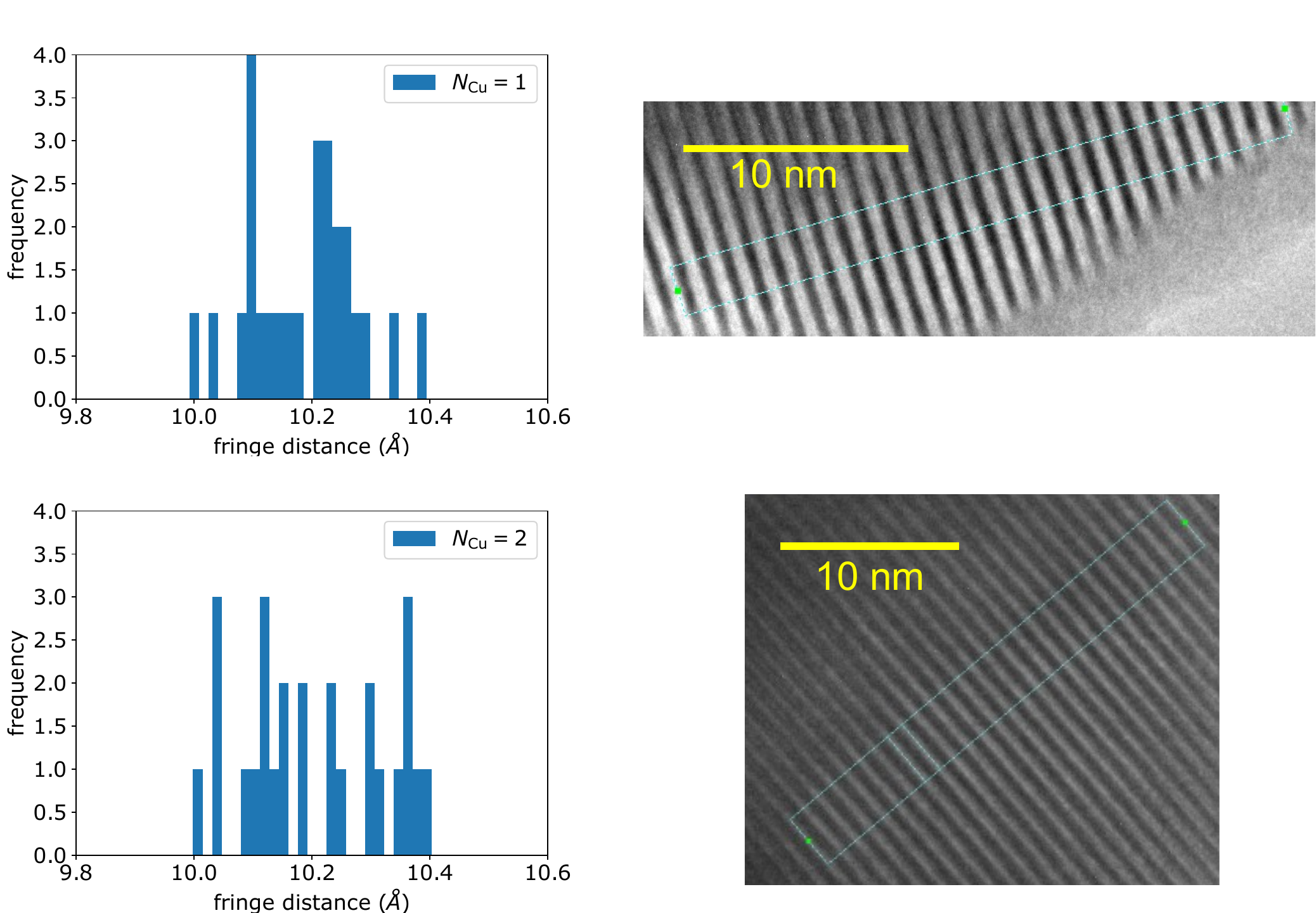} 
    \caption{ \edfigurelabel{fig:TEM} Histogram of representative samples from the Cu-deficient ($N_\mathrm{Cu}=1$, above) and near-stoichiometric ($N_\mathrm{Cu}=2$, below) TEM patterns. Note that each pixel represents 0.44 \AA-distance, and the calculated fringe distribution all fall within this pixel.} 
\end{figure} 


\section{Magnetization and Susceptibility} 
\label{Mag_and_Susc}
The volume magnetization was deduced from the total moment $\mu_\mathrm{tot}$ from SQUID measurements by taking into account the mass $m$ of the samples, and using theoretical density $\rho_\mathrm{th} = 7.48$ g cm$^{-3}$, to compute $M = \mu_\mathrm{tot}/V = \mu_\mathrm{tot} \rho_\mathrm{th}/m$. In magnetization data, we computed the internal field to correct for the effects of the demagnetization factor, 
 \begin{equation} 
     H_{\mathrm{int},i} = H_{a,i} - 4\pi N_i M_i 
 \end{equation} 
where $H_{a,i}$ is the applied field in the direction $i$, $M_i$ is the (volume) magnetization, and $N_i$ is the demagnetization factor with $0\leq N_i \leq 1$. For diamagnetic samples in a rectangular prism geometry, the demagnetization factor could be estimated with \citesec{Prozorov}
 \begin{equation}
     N \approx \frac{4AB}{4AB+3C(A+B)}
 \end{equation}
where $A,B$ ($C$) are the lengths of the sides perpendicular (parallel) to the applied field. Anisotropic magnetization measurements on an optimized sample $(N_\mathrm{Cu} = 2$) were performed on a nearly-rectangular-prism sample with approximate dimensions 3.2 $\times$ 1.52 $\times$ 0.84 mm$^3$ and mass $27.54$ mg. From this, we estimate a demagnetization  factor $N_a \approx 0.18$ for fields along the $a$-axis and $N_c \approx 0.62$ for fields along the $c$-axis. In the $\mu$SR measurement, our collection of co-aligned samples formed the shape of a rectangular prism with effective dimensions 15 $\times$ 12 $\times$ 1.4 mm$^3$. Here the magnetic field was always applied parallel to the thinnest dimension, yielding a demagnetization correction factor of about $N\approx 0.86$. 

Quantities like the superconducting transition in susceptibility were estimated by fitting to simple models of a diamagnetic response assuming a Gaussian distribution. Supposing the susceptibility is exactly $4\pi\chi =-1$ below $T_c$ and exactly $4\pi \chi = 0 $ above $T_c$, we get a step-like transition at $T_c$. For $T_c'$ distributed in a Gaussian distribution about some mean $T_c$, this becomes
 \begin{equation} \label{eq:chi_fit}
    4\pi \chi (T) = \int_{-\infty}^\infty \dd T_c' [\Theta(T-T_c') -1] \frac{e^{-(T_c'-T_c)^2/2\sigma^2}}{\sqrt{2\pi \sigma^2}} = \frac{1}{2} \left[ \mathrm{erf} \left( \frac{T-T_c}{\sqrt{2\sigma^2}} \right) - 1 \right] \equiv B_{\sigma}(T,T_c)
 \end{equation} 
For the ambient-pressure measurements, the susceptibility for all samples measured showed sharp transitions and no broad tails at low temperatures, with all samples having a saturated susceptibility by 0.4 K. In all cases $\sigma$, the width of the superconducting transition, was used to estimate the error bar on $T_c$, rather than the error reported by the fit routine. The susceptibility for various values of $N_\mathrm{Cu}$ is shown in Extended Fig.~\ref{fig:susceptibility}a.  

For the high-pressure measurements above 1.7 GPa, the AC data were fit using a modified form due to the double-peak nature of the transition. We model this double-peak feature coming from the successive transition of two gaps opening at $T_c$, for the sake of extracting $T_c(p)$ and $T^*$: 
 \begin{equation} \label{eq:hydrostatic_fit} 
    4\pi \chi'(T,p) = A\left[ fB_{\sigma_1}(T,T_{c}(p)) + (1-f)B_{\sigma_2}(T,T^*(p)) \right] + C 
 \end{equation} 
where $A$ is the voltage amplitude, $f$ is the fraction of each component, $\sigma_1$ and $\sigma_2$ are the widths of the transitions, and $C$ is a constant voltage offset. 

Ambient-pressure susceptibility data are shown in Extended Fig.~\ref{fig:susceptibility}b, highlighting the zero-field cooled (ZFCW) and field-cooled (FCW) curves for the optimized sample (measured upon warming), using $4\pi \chi_0 = 4\pi \chi_v (1-N_i)$ where $\chi_v$ is the molar susceptibility calculated with the nominal applied field value. Note that $4\pi \chi$ is greater than $-1$, due to the large relative error suggesting the applied field might have been slightly larger than the reported 2 Oe. While the FC volume susceptibility $4\pi\chi_v$ is an indication of the Meissner fraction, errors in field calibration or density estimations resulted in inferred Meissner fractions greater than $100\%$. However, the demagnetization-corrected magnetization versus applied field demonstrate the $4\pi \chi_v = -1$ relation in Fig.~\ref{fig:fig3}c,d. 

\begin{figure}
    \centering
    \includegraphics[width=\linewidth]{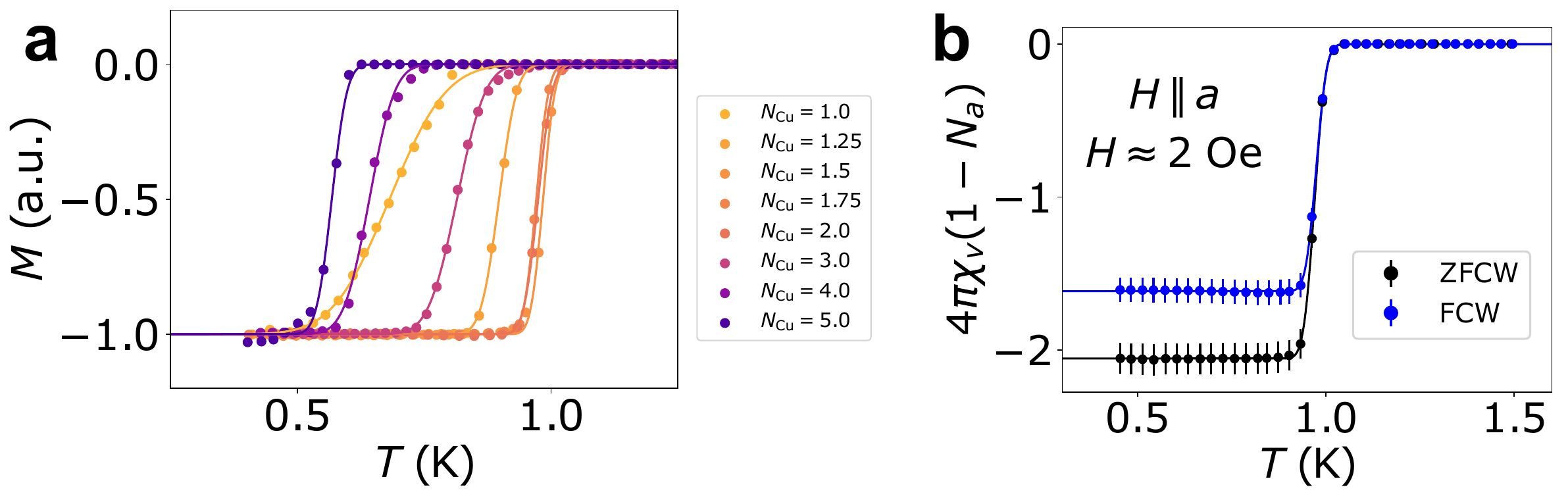}
    \caption{ \edfigurelabel{fig:susceptibility} (a) The susceptibility curves for \ch{LaCuSb2} (normalized to the saturation value) for various values of $N_\mathrm{Cu}$. The points are the data and the solid lines are the fits to Eq.(~\ref{eq:chi_fit}). (b) Zero-field cooled and field-cooled susceptibility, measured upon warming. Solid lines are fits to Eq.(~\ref{eq:chi_fit}). }
\end{figure}

To determine the critical field from $c$-axis magnetization data $H_\mathrm{int,c}(T)$, we find the field that results in a discontinuous transition into the normal state ($M\neq 0$ to $M=0$). This was done by fitting the $H_\mathrm{int}$ data versus $4\pi M_c$ to a constant value, for each fixed temperature. For magnetization along the $a$-axis, we define the critical field $H_{c1}$ to be the inflection point of the magnetization. Numerically, this was determined by taking the derivative data $f = \dd (4\pi M)/\dd H$, and fitting it near $H_{c1}$ to the piecewise function 
 \begin{equation}
     f(H,H_0) = 
     \begin{cases} 
      -1 & H\leq H_0 \\
      a(H-H_0)-1 & H>H_0
     \end{cases}
 \end{equation}
The critical field $H_{c1}$ was defined as the field where $4\pi M$ changes slope, or $f(H_{c1},H_0) = 0$, thus
 \begin{equation}
     H_{c1} = \frac{aH_0 + 1}{a} 
 \end{equation}
Furthermore, $H_{c2}$ was determined by fitting $4\pi M$ as a linear function at high fields (for small values of $4\pi M$), 
 \begin{equation}
     4\pi M \approx bH + c \qquad (4\pi M\approx 0)
 \end{equation} 
The critical field was defined such that $4\pi M(H_{c2}) = 0$, or $H_{c2} = -c/b$ from the fitted parameters. In all of these fits, the uncertainty was obtained through error propagation. 

We can finally extract the zero-temperature critical field by fitting the data to conventional models for the temperature dependence: 
 \begin{gather} \label{eq:critical_fields}
     H_{c1}(T) = H_{c1}(0) \cdot (1-t^2) \\
     H_{c2}(T) = H_{c2}(0) \cdot \frac{1-t^2}{1+t^2}  
 \end{gather}
where $t=T/T_c$ is the reduced temperature. This was done for the data in magnetization, specific heat, and $\mu$SR where we assumed only the $a$-axis oriented samples have a critical field $H_{c2}$, while the remainder of the data were fit for $H_{c1}$.

\section{Transport} \label{SI_Transport}
The reason why superconductivity exists in some samples and not in others with differing copper content is an important question. Given the complicated band structure with various bands crossing the Fermi level, it is not out of the question that tuning the stoichiometry will affect disorder scattering. Indeed, the resistivity of the optimized sample is about an order of magnitude lower than that reported for \ch{LaCuSb2} in the literature \cite{Chamorro, Muro}. As seen in Extended Fig.\ref{fig:resistivity_comparison}a, the residual resistivity at low temperatures is $\rho_{0a} = 1.883(15) \; \mu\Omega$-cm with a residual resistivity ratio (RRR) of about 14, whereas previous samples have been in the $~\sim 1$ m$\Omega$-cm range. Furthermore, the low-temperature residual resistivity is the lowest in the optimized sample $N_\mathrm{Cu}=2$, whereas it is larger for other samples near the endpoints of the superconducting dome. Importantly, the large linear magnetoresistance seen in Extended Fig.~\ref{fig:resistivity_comparison}b appears resilient in \ch{LaCuSb2}, as it is found in the normal state of the superconducting samples as well as in the samples from Chamorro \textit{et al.} \cite{Chamorro}. This suggests that the Dirac electrons still play an important role at low temperatures in our samples. 

\begin{figure}
    \centering
    \includegraphics[width=\linewidth]{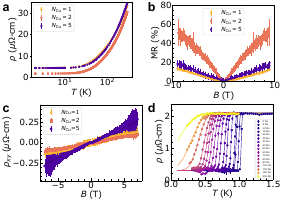}
    \caption{ \edfigurelabel{fig:resistivity_comparison} Standard PPMS AC transport on samples grown with $N_\mathrm{Cu}=1,2,5$. (a) The in-plane, zero-field resistivity highlights that the optimized $N_\mathrm{Cu}=2$ superconducting sample has the lowest residual resistivity compared to the end members of the superconducting dome. (b) Symmetrized MR in superconducting samples at 2 K, showing significant linearity in all samples. The largest MR in these samples occurs for $N_\mathrm{Cu}=2$, with highest superconducting $T_c$. (c) Antisymmetrized Hall effect data on same samples in the normal state at 2 K. The fits to the single-band Hall model for $N_\mathrm{Cu}=2$ yields $4.15(3)\times 10^{22}$ cm$^{-1}$. (d) DR resistivity as a function of temperature and applied field $H\parallel c$. Solid lines are the fits to Eq.(~\ref{eq:erf2}), assuming a constant offset.} 
\end{figure}

Besides effects of disorder, varying the copper concentration can also affect the chemical potential, and thus the carrier density. To study this, Hall effect measurements were taken using a five-probe configuration with AC Transport in the PPMS, on a sample cut from the same crystal as used in other measurements. The lead separation was $0.38$ mm and the sample thickness was $0.15$ mm. The applied current was 40 mA with frequency 103 Hz. Symmetrization difficulties  resulted in apparent non-linearity. Regardless, the Hall coefficient gave a carrier density of approximately $4.16(2)\times 10^{22}$ cm$^{-1}$. Interestingly, the sign of the charge carriers in the sample with $N_{\mathrm{Cu}} = 2$ was positive, while the sign for samples from Chamorro \textit{et al.} \cite{Chamorro} was negative. This suggests the change in stoichiometry also results in a change in the Fermi level, which can result in different contributions from electron or hole charge carriers. In our work, there is also an apparent trend that the end members of the superconducting dome have a smaller carrier density than the optimized sample, for example seen in Extended Fig.~\ref{fig:resistivity_comparison}c. Off-stoichiometry  not only affects the sample quality but also the electronic structure, and may correlate with the presence of superconductivity at low temperatures. However, the carrier densities across the superconducting dome are within an order of magnitude of each other, so the density of states is likely to change in proportion to the small changes in the Copper content $x_\mathrm{Cu}$.

We also studied the resistivity in the superconducting state for the optimized $N_\mathrm{Cu}=2$ sample. The DR-temperature resistivity data in applied fields are shown in Fig.~\ref{fig:resistivity_comparison}d. Like the susceptibility, the temperature dependence of the resistivity was fit assuming Gaussian broadening:
 \begin{equation} \label{eq:erf2}
     \rho(T) = \frac{1}{2} \left[ \mathrm{erf} \left( \frac{T-T_c}{\sqrt{2\sigma^2}} \right) + 1 \right] \rho_{0a}
 \end{equation}
with $\rho_{0a}$ the low-temperature residual resistivity. The resistivity was measured for applied magnetic field along the $c$-axis and currents along the $a$-axis, varying temperature at fixed fields. In these measurements we did not reach a zero-resistance state, but the temperature of the midpoint $T_c$ found in the resistivity drop is consistent with $T_c$ from other thermodynamic measurements, in low fields $H<25$ Oe. At fields higher than $H_{c1}$, the width $\sigma$ became exceedingly large and the resistance did not saturate at low temperatures, so it was difficult to estimate $T_c$ and $\sigma$. This also suggests that the surviving superconducting state at these higher fields is due to percolation or filamentary superconductivity, and is not necessarily a bulk response. For this reason, we did not fit the resistivity data to extract critical field(s) in Extended Fig.\ref{fig:phase_diagram}.

Overall, resistivity data for various samples highlights the importance of the copper stoichiometry. While the susceptibility data indicate changes in the superconductivity, we find concomitant changes in the resistivity and Hall effect that suggest these phenomena are related. However, despite changes in defect density, Fermi level, and carrier densities, we still find large linear magnetoresistance in all samples suggestive of the presence of Dirac fermions.

\section{Specific Heat} \label{SISpecificHeat} 
To extract the transition temperatures from specific heat, we first fit the data $c_p/T$ to a spline near the transition temperature. The derivative of this spliced data was taken, and the resulting derivative data were fit to a Gaussian function. The transition temperature was then reported as the midpoint and standard deviation of this fit function, at various fields, for use in the phase diagram. 

In the normal state above 1 K, we extrapolate the Sommerfeld coefficient and phonon contribution $c_p = \gamma_n T + \beta_3 T^3$ to zero temperature and to lowest order. This yields a Sommerfeld coefficient $\gamma_n = 4.78(1)$ mJ/mol-K$^2$ and a phonon contribution $\beta_3 = 0.571(2)$ mJ/mol-K$^4$. With $d=4$ atoms per formula unit \ch{LaCuSb2}, the corresponding Debye temperature $\Theta_D = (12\pi^4 Rd/5\beta_3)^{1/3} = 238.8(3)$ K. This is consistent with previously reported values \cite{Muro}. 

In the isotropic free-electron model, the molar specific heat Sommerfeld coefficient $\gamma_n$ in the normal state can be used to determine the specific heat effective mass of the charge carriers \citesec{Ashcroft}, 
 \begin{equation} \label{eq:gamma}
     \gamma_n = (V_\mathrm{fu} N_A) \gamma_{nV} = \frac{\tfrac{1}{2} \pi^2 V_0 Rk_B n}{(\hbar k_F)^2} m^*
 \end{equation}
where $\gamma_{nV}$ is the volume specific heat Sommerfeld constant, and $V_0$ and $V_\mathrm{fu} = \frac{1}{2} V_0$ are the volumes of the unit cell and one formula unit, respectively. Using the measured Sommerfeld constant, along with the carrier density derived from the Hall effect, we find a specific heat effective mass of $m^* = 1.44(1)m_e$. This is to be contrasted with the low effective in-plane masses of charge carriers in our previous report \cite{Chamorro}, likely due to the anisotropy of the Fermi surface. 

From specific heat one can obtain the electron-phonon coupling parameter $\lambda_{e-p}$, given by 
 \begin{equation} \label{eq:coupling_parameter}
     \lambda_{e-p} = \frac{1.04+\mu^* \ln\left(\dfrac{\Theta_D}{1.45T_c}\right)}{\left(1-0.62\mu^* \right)\ln\left(\dfrac{\Theta_D}{1.45T_c}\right)-1.04}
 \end{equation}
where $\mu^* \approx 0.13$ for intermetallic superconductors. We find a value of $\lambda_{e-p} \approx 0.466$, which is in contrast to the value assumed in a previous theoretical work finding $\lambda_{e-p} \approx 0$ \citesec{Ruszala}. However, the free-electron value $\gamma_b \approx 2.85$ mJ/mol-K$^2$ and the experimental value differ by the enhancement factor $\lambda_{e-p} = \gamma_n/\gamma_b - 1 \approx 0.678$, in reasonable agreement with the calculated value in Eq. (\ref{eq:coupling_parameter}).

The main contribution to the total specific heat at lowest temperatures is from the nuclear Schottky anomaly. This is ascribed to the interaction of nuclear quadrupolar moments with local electric field gradients at the nucleus (see SI~\ref{SISpecificHeat} for more details on the corresponding modeling). This contribution is dependent on both the spinful nuclear moments, the point group symmetry of the ions involved, and is to be expected for $T<100$~mK in low symmetry solids containing Cu \citesec{Caspary} and/or Sb  \citesec{Ortiz}. The anomaly also grows with magnetic field, as seen in other Sb-based superconductors \citesec{Aoki}. To extract the electronic specific heat, we model the total contribution to the specific heat for $T<0.35$~K by the equation
 \begin{equation} \label{eq:SH_fit}
    c_p(T,H) = c(H) e^{-\Delta(H)/k_B T} + A(H)/T^2 + \beta_3 T^3
 \end{equation}
where $c(H)$ and $\Delta(H)$ are the phenomenological parameters used to model the activated behavior associated with the knee-like feature; $A(H)$ is related to the quadrupole coupling and increases in applied magnetic fields as the nuclear spin states undergo additional Zeeman splitting. In this way, we simultaneously fit the different contributions to separate out the Schottky anomaly (dominant at low temperature) and phonons (dominant at high temperature) from the electronic contributions. 

The entropy of the electronic charge carriers is calculated as 
 \begin{equation}
    \frac{\Delta S_\mathrm{el}(T)}{\gamma_n T} = \frac{1}{T} \int_0^T \frac{c_\mathrm{el}(T')}{\gamma_n T'} \dd T' 
 \end{equation} 
For free electrons, $\Delta S_\mathrm{el}(T)/\gamma_n T = 1$ at all temperatures. In a superconducting sample, $\Delta S_\mathrm{el}(T)/\gamma_n T = 1$ for $T>T_c$ in the normal state, and there is an entropy balance at $T_c$ such that $\Delta S_\mathrm{el}(T_c)/\gamma_n T_c = 1$. For $T<T_c$ in the superconducting state, $\Delta S_\mathrm{el}(T_c)/\gamma_n T_c<1$ and decreases to zero as $T\to 0$. Extended Fig.~\ref{fig:entropy} shows the calculated entropy, which is close to 1 at $T_c$ and above. This indicates the sample is a bulk superconductor and that the sharp drop in the electronic specific heat at $T^*$ is intrinsic. Note that the sample coupling reported by the PPMS decreases at decreasing temperatures, due to the combination of the exponentially-activated thermal conductivity and the large sample mass required for a good measurement signal. However, this sudden drop near $T^*$ is necessary to obey entropy balance at $T_c$, where $\Delta S(T_c) = \gamma_n T_c$. The slight overshoot of the entropy $\Delta S_\mathrm{el}/\gamma_n T_c > 1$ is likely a result of the uncertainty associated with subtracting the large nuclear Schottky anomaly when isolating the electronic contribution to the specific heat capacity. Furthermore, around phase transitions the specific heat can be difficult to extract $C_p$ precisely, adding further error to calculating the entropy. These limit the rigor possible when fitting to multi-gap models, as these BCS and self-consistent models strictly obey entropy balance. 

\begin{figure}
    \centering
    \includegraphics[width=\linewidth]{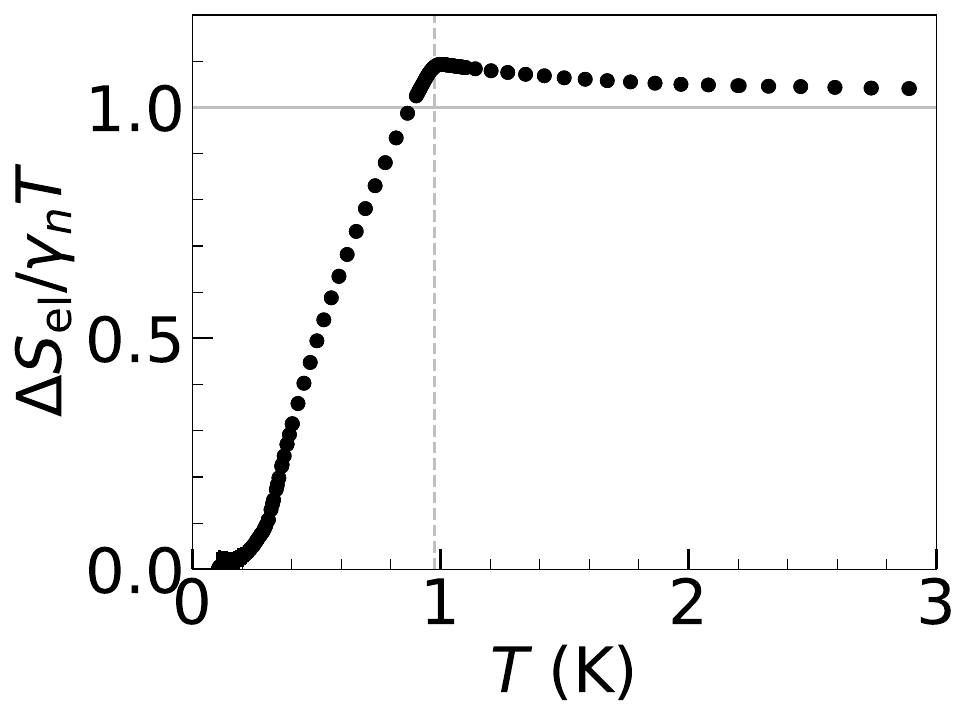}
    \caption{\edfigurelabel{fig:entropy} Computed entropy from the electronic specific heat. We note $\Delta S/\gamma_n T$ is close to 1 (solid line) at and above $T_c$ (dashed line). } 
\end{figure}

\section{Zero-field muon spin rotation} 
The existence of Dirac fermions in \ch{LaCuSb2}, coupled with any unconventional time-reversal symmetry (TRS) breaking, would make \ch{LaCuSb2} a prime material candidate in the search for monopole superconductivity. TRS breaking is not \textit{a priori} expected in \ch{LaCuSb2}, since the nominal crystal structure P4/nmm is centrosymmetric, and all ions in \ch{LaCuSb2} are nonmagnetic. To determine whether TRS is broken below $T_c$ in \ch{LaCuSb2}, we used $\mu$SR in zero field as a way to search for inhomogeneous magnetic fields. Extended Fig.~\ref{fig:ZF_asy} shows the asymmetry $A(t)$ as a function of time for temperatures above and below $T_c$. At all temperatures, we fit the asymmetry to the equation 
 \begin{equation} \label{eq:KT-ZF} 
    A(t) = A_0 [F \cdot G_{KT}(t) e^{-\Lambda t} + (1-F) e^{-\lambda_\mathrm{bg} t}]
 \end{equation} 
where $G_{KT}(t)$ is the Kubo-Toyabe function due to random fields from nuclear moments; $\Lambda$ is the temperature-dependent relaxation rate; and $F$ is the fraction of muons that stop in \ch{LaCuSb2} as opposed to the silver mounting plate. In exotic superconductors that break TRS, $\Lambda(T)$ increases with decreasing temperature due to larger field inhomogeneity that often appears below $T_c$, such as from domains related by TRS. Simultaneous fits of our high- and low-temperature ZF spectrum reveal a relaxation rate $\Lambda(T)$ that is essentially constant with temperature (within error), where $\Lambda = 9(4)\times 10^{-3}$ $\mu$s$^{-1}$ at 1.293(3) K and $\Lambda = 11(4)\times 10^{-3}$ $\mu$s$^{-1}$ at 0.017(1) K. That is, the maximum field size that could exist consistent with these data is $\Delta \Lambda/\gamma_\mu = 0.02(7)$ Oe.  
\begin{figure}
    \centering
    \includegraphics[width=\textwidth]{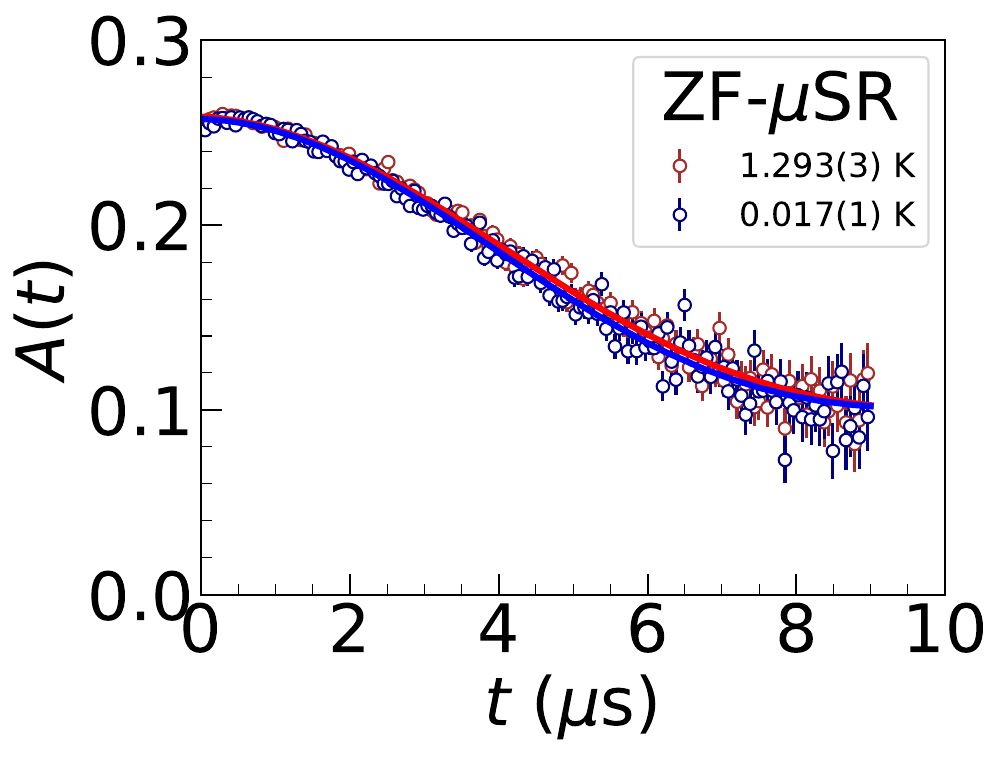}
    \caption{ \edfigurelabel{fig:ZF_asy} Zero-field $\mu$SR data for \ch{LaCuSb2}, showing the asymmetry as a function of time at temperatures below (blue) and above (red) $T_c$. The fits to the data produce relaxation rates $\Lambda(T)$ that are within error of each other.} 
\end{figure}

\section{Vortices and transverse-field muon spin rotation} \label{ssec:TF_muSR}
For a Type-II superconductor, the relaxation rate $\sigma$ is related to the superconducting relaxation rate by $\sigma = \sqrt{ \sigma_\mathrm{SC}^2 + \sigma_\mathrm{n}^2 }$, where $\sigma_{\textrm{n}}$ is the temperature independent nuclear spin induced relaxation rate. We estimate the London penetration depth from $\sigma_\mathrm{SC}$ as follows. Firstly, for $H\parallel a$, muons probe the London penetration depth $\lambda_{ac} = \sqrt{\lambda_a \lambda_c}$ due to superconducting currents flowing in the $ac$ plane \citesec{Liarte}. Assuming that the field ratio $b\equiv H_\mathrm{app}/H_{c2}$ satisfies $0.13/\kappa^2 \ll b \ll 1$, where $\kappa$ is the GL parameter, the London penetration depth and superconducting relaxation rate are related by \citesec{Brandt} 
 \begin{equation} \label{eq:London_relaxation}
     \sigma_\mathrm{SC} \approx \frac{0.0609 \gamma_\mu \Phi_0}{\lambda^2}, 
 \end{equation} 
where $\Phi_0 \approx 2.067\times 10^{-15}$ Wb is the magnetic flux quantum. With an applied field of $H_\mathrm{app} = 40$ Oe, we were above $H_{c1} = 32(1)$ Oe to set us firmly in the mixed phase, and such that $b\approx 40$ Oe$/172$ Oe $\approx 0.23$ satisfies the condition that $0.13/\kappa^2 \ll b \ll 1$, assuming $\kappa_a \sim 1$. 

In this way, we were able to estimate the geometric mean of the zero-temperature anisotropic London penetration depth, $\lambda_{ac}(0) \approx 408(2)$ nm. We note that more accurate estimations of $\lambda_{ac}(0)$ will involve functions of $b$ and $\kappa$, which we do not extract directly from the $\mu$SR data. We can extract the superfluid density $\rho(T)=\lambda^2(0)/\lambda^2(T)$ independent of the assumptions made in Eq. (\ref{eq:London_relaxation}) if we assume the carrier effective mass is constant.

\section{Tight-binding model for Fermi surfaces of topological bands} \label{SITB}
\begin{figure}
    \centering
    \includegraphics[width=0.7\textwidth]{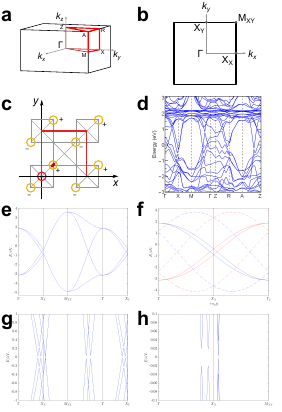}
    \caption{ \edfigurelabel{fig:TightBinding} Band structure and tight-binding analysis of \ch{LaCuSb2}. (a) Brillouin zone, high symmetry points and high-symmetry paths for space group $P4/nmm$.  (b) Brillouin zone in the $k_z=0$ plane.  (c) Model unit cell of $\ch{Sb}$ square net. The inversion center is indicated by the red circle in between the $\ch{Sb}$ sites, and the two-fold rotational symmetry along the $z$-axis is centered on $\ch{Sb}$ sites.  (d) Band structure calculated by Density Functional Theory.  (e) Tight binding model band structure with parameters $t_1 = 3.375$, $t'_1 = 0.875$, $t_2=0.125$, $t'_2=0.125$, and $\epsilon_p-\mu = -0.625$.  (f) Band cuts along $\mathbf{G}_x= 2\pi/a \hat{x}$ from $\Gamma$ (solid curves) and $\Gamma+\pi/2 \hat{y}$ (dashed curves). Bands with $M_z$ eigenvalue $-1$ are in red and $+1$ are in blue.  (g) BdG Spectrum. Band parameters are  as before. Typical values of $\Delta_1=20\text{meV}$ and $\Delta_2=20\text{meV}$ are used.  (h) Plots of the band structure focusing on the low-energy regions. The pairing amplitudes are set large compared to the experimental value for demonstration purpose.}
\end{figure}

\ch{LaCuSb2} comprises multiple Fermi surfaces. Besides several small pockets around the $\Gamma$-point, there are a pair of large quasi-$2$D diamond-shaped Fermi surfaces. Along with the pockets around the $X$-point, these arise from the topological bands with Dirac nodal lines protected by the nonsymmorphic symmetries of the space group. To better demonstrate the band topology and study the consequences for superconductivity, we construct an eight-band tight-binding model that captures essential features of the Fermi surfaces of topological bands. 

According to the first principles calculations, the topological bands mainly consists of $p_{x,y}$-orbitals of $\ch{Sb}$, which allows us to focus on a $2$D $\ch{Sb}$ square net layer. The $\ch{Sb}$ square net is geometrically a square lattice where $\ch{Sb}$ atoms occupy the lattice sites. However, the unit cell of the square net (defined by the base vectors $\mathbf{a}$ and $\mathbf{b}$ of $\ch{LaCuSb2}$) is the doubled unit cell of the square lattice. Therefore, in each unit cell of $\ch{Sb}$ square net, there are two $\ch{Sb}$ atoms. These two sites per unit cell can be artificially distinguished by slightly displacing the atoms along $\pm{z}$-directions.
It should be emphasized that, in reality, the $\ch{Sb}$ atoms forming the square net are geometrically all in a plane, and the difference comes from the chemical environment of the neighbouring layers above and below.
Nevertheless, the artificial displacements reduce the accidental symmetries of a square lattice to the true symmetries of the layer group P4/nmm inherited from the space group bearing the same name.

Extended Fig.~\ref{fig:TightBinding}a,b show the high-symmetry points and paths in the Brillouin zone for space group $P4/nmm$, along with those at $k_z=0$, respectively. Extended Fig.~\ref{fig:TightBinding}c shows the unit cell of $\ch{Sb}$ square net. The orange circles represent the $\ch{Sb}$ atoms. The inversion center is indicated by the red circle in between the $\ch{Sb}$ sites, and the two-fold rotational symmetry along $z$-axis is centered on $\ch{Sb}$ sites. 

After putting $p_{x,y}$-orbitals on each $\ch{Sb}$ site, and labeling the two sublattices by $r = \{A,B\}$, we now describe our tight-binding model. The nearest neighbour hoppings and second nearest neighbour hoppings are parameterized by Slater-Koster type parameters. Notice that, for nearest neighbour hoppings, the hoppings are along diagonals, so
the orbital basis are rotated by $45^\circ$ as
\begin{equation}\label{eq:TBModel_BasisRotation}
\begin{split}
\begin{pmatrix}c_{r,p_\xi} \\ c_{r,p_\eta}\end{pmatrix}
= \frac{1}{\sqrt{2}} 
\begin{pmatrix}
1 & 1\\
-1 & 1
\end{pmatrix}
\begin{pmatrix}c_{r,p_x} \\ c_{r,p_y}\end{pmatrix}.
\end{split}
\end{equation}
The Hamiltonian for the nearest neighbour hoppings can be written as
\begin{equation}\label{eq:TBModel_TBBandnnhopping}
\begin{split}
\mathcal{H}_{nn} 
=& \frac{1}{2} \sum_{\mathbf{R},\bm{\delta},\alpha} 
t_{1;\bm{\delta},\alpha}
c_{\mathbf{R}+\bm{\delta}, B, p_{\alpha}}^\dagger 
c_{\mathbf{R}, A, p_{\alpha}}
+ \text{h.c.}\\
=&\sum_{\mathbf{k}} 
\psi_{\mathbf{k}, r, p_{\alpha}}^\dagger 
\begin{pmatrix}
h_\xi(\mathbf{k}) & 0\\
0 & h_\eta(\mathbf{k})
\end{pmatrix}
\psi_{\mathbf{k}, r, p_{\alpha}},\\
t_{1;\bm{\delta},\alpha} =& \begin{cases}
+t_1, & \delta = \alpha,\\
-t'_1, & \delta \neq \alpha;
\end{cases}\\
\psi_{\mathbf{k}, r, p_{\alpha}}^\dagger 
=& \begin{pmatrix}c_{\mathbf{k},A, p_{\xi}}^\dagger & c_{\mathbf{k}, B, p_{\xi}}^\dagger &
c_{\mathbf{k},A, p_{\eta}}^\dagger & c_{\mathbf{k}, B, p_{\eta}} ^\dagger \end{pmatrix},\\
h_\xi(\mathbf{k}) = &\left[+t_1 \cos\left(\frac{k_x a}{2} + \frac{k_ya}{2}\right) - t'_1\cos\left(-\frac{k_x a}{2} + \frac{k_ya}{2}\right)\right] \tau^{(r)}_1 ,\\
h_\eta(\mathbf{k}) = & \left[ - t'_1\cos\left(\frac{k_x a}{2} + \frac{k_ya}{2}  \right) +t_1 \cos\left(-\frac{k_x a}{2} + \frac{k_ya}{2}\right) \right] \tau^{(r)}_1,
\end{split}
\end{equation}
where the sum of $\mathbf{R}$ runs through all the unit cells, and $\bm{\delta}$ points towards the nearest neighbouring sites. The matrices $\tau_i^{(r)}$ are Pauli matrices with superscript specifying the physical degrees of freedom the Pauli matrices act upon. Here $(r)$ indicates the sublattice index. We will use $\tau^{(\sigma)}_i$ for spins and $\tau^{(\Delta)}_i$ for particle-hole doubling in the Bogoliubov-de Gennes formalism of superconductivity.  

The Hamiltonian for the second nearest neighbour hoppings can be written in a similar fashion as
\begin{equation}\label{eq:TBModel_TBBandsnnhopping}
\begin{split}
\mathcal{H}_{2nn} 
=& \frac{1}{2} \sum_{\mathbf{R},\bm{\delta}',\alpha} 
t_{2;\bm{\delta}',\alpha} \left[
c_{\mathbf{R}+\bm{\delta}', A, p_{\alpha}}^\dagger 
c_{\mathbf{R}, A, p_{\alpha}} + c_{\mathbf{R}+\bm{\delta}', B, p_{\alpha}}^\dagger 
c_{\mathbf{R}, B, p_{\alpha}} \right]\\
=&\sum_{\mathbf{k}} 
\psi_{\mathbf{k}, r, p_{\alpha}}^\dagger 
\begin{pmatrix}
h_x(\mathbf{k}) & 0\\
0 & h_y(\mathbf{k})
\end{pmatrix}
\psi_{\mathbf{k}, r, p_{\alpha}},\\
t_{2;\bm{\delta}',\alpha} =& \begin{cases}
+t_2, & \delta' = \alpha,\\
-t'_2, & \delta' \neq \alpha;
\end{cases}\\
\psi_{\mathbf{k}, r, p_{\alpha}}^\dagger 
=& \begin{pmatrix}c_{\mathbf{k},A, p_{x}}^\dagger & c_{\mathbf{k}, B, p_{x}}^\dagger &
c_{\mathbf{k},A, p_{y}}^\dagger & c_{\mathbf{k}, B, p_{y}} ^\dagger \end{pmatrix},\\
h_x(\mathbf{k}) = &\left[+t_2 \cos(k_xa) -t'_2 \cos(k_ya) \right] \tau_0^{(r)},\\
h_y(\mathbf{k}) = & \left[ - t'_2\cos(k_xa ) +t_2 \cos(k_ya) \right] \tau^{(r)}_0,
\end{split}
\end{equation}
where $\bm{\delta}'$ points towards the second nearest neighbour sites.

We also include the orbital energies and the chemical potential for the sake of describing doping and discussing superconductivity. These are simply diagonal terms
\begin{equation}\label{eq:TBModel_TBBandChemPot}
\begin{split}
\mathcal{H}_{\mu} 
=& \sum_{\mathbf{R},r,\alpha} 
(\epsilon_{p} - \mu )\left[
c_{\mathbf{R}, r, p_{\alpha}}^\dagger 
c_{\mathbf{R}, r, p_{\alpha}} \right]\\
=&\sum_{\mathbf{k}} 
\psi_{\mathbf{k}, r, p_{\alpha}}^\dagger 
\begin{pmatrix}
(\epsilon_{p}-\mu)\tau_0^{(r)} & 0\\
0 & (\epsilon_{p}-\mu)\tau_0^{(r)} 
\end{pmatrix}
\psi_{\mathbf{k}, r, p_{\alpha}},
\end{split}
\end{equation}
where we assume the orbital energies $\epsilon_{r,p_\alpha} = \epsilon_{p}$ are the same for both sublattices required by the crystal symmetry (e.g., the inversion symmetry).

To determine the tight-binding parameters, we fit our model band structure to the first principles calculations. The full band structure is shown in Extended Fig.~\ref{fig:TightBinding}d. We find $t_1 = 3.375$, $t'_1 = 0.875$, $t_2=0.125$, $t'_2=0.125$, and $\epsilon_p-\mu = -0.625$ all in  unit of $\text{eV}$. Extended Fig.~\ref{fig:TightBinding}e shows our model bands that give rise to the Dirac nodal lines.

The band structure in Extended Fig.~\ref{fig:TightBinding}e exhibits several band crossings near the $X$-point. Bands that are degenerate at the $X$-point remain doubly degenerate along $X$-$M$, which gives rise to a dispersive Dirac nodal line. We also note that the crossings between $\Gamma$-$X$ and $\Gamma$-$M$ have the same origin. They are, in fact, part of the diamond-shaped Dirac nodal line intersecting with the high symmetry planes. The Dirac nodal lines are protected by the non-symmorphic crystal symmetry, as we will demonstrate below.

The key symmetry element that protects the nodal lines is the glide mirror plane $g=\{R=M_z|\mathbf{t}= [1/2,1/2,0]\}$. For any $\mathbf{k}$ in $k_z=0$ plane, 
$g\mathbf{k} = \mathbf{k}$,
therefore, we may choose the Bloch waves to be eigenstates of $g$ as well, i.e. $g|u_\mathbf{n,k} \rangle = \lambda_{n,\mathbf{k}}|u_\mathbf{n,k} \rangle, \lambda_{n,\mathbf{k}}= \lambda_n \exp(\mathrm{i}\mathbf{k} \cdot \mathbf{t})$.
Since $R^2=1$, $\lambda_n = \pm1$. Extended Fig.~\ref{fig:TightBinding}f shows band cuts along $\mathbf{G}_x= 2\pi/a \hat{x}$.  The solid curves are bands cut through the $\Gamma$-point, and the dashed curves are bands cut through $\Gamma+\pi/2 \hat{y}$. Band eigenvalues are calculated according to our tight-binding Hamiltonian. Bands with eigenvalue $\lambda=-1$ are in red and $\lambda=+1$ in blue. Furthermore, the time-reversal and the inversion symmetries ensure two-fold degeneracy of each band due to spins. Therefore, the symmetry protected crossings are Dirac nodes.

It bears emphasizing that the nodes on BZ boundaries and those within the BZ are of different types. The TR symmetry $\Theta$ together with the non-symmorphic symmetry $g$ protects the nodal line along $X$-$M$, which can be understood as Kramers degeneracy with respect to the antiunitary operator $\tilde{\Theta} = g \Theta$, with $\tilde{\Theta}^2=-1$. Therefore, these crossings, located at BZ boundaries, are like type-II Dirac node.

Regarding four bands as two pairs of intertwined bands crossing at BZ boundaries, the band crossings between the pairs are also protected by $g$. However, these crossings are less robust compared to the previous case, in the sense that when the pairs of bands are deformed, the inter-pair crossings can move and even annihilate pairwise. Extended Fig.~\ref{fig:TightBinding}f shows two cuts along $\mathbf{G}_x= 2\pi/a \hat{x}$ through the $\Gamma$-point and $\Gamma+\pi/2 \hat{y}$.  Note that the inter-pair crossing moves towards $k_x=0$. These crossings can be finally annihilated and give rise to a closed diamond-shaped nodal line inside BZ. Therefore, unlike the crossings at BZ boundaries, the inter-pair crossings are like type-I Dirac nodes. 

In the presence of the spin-orbit couplings (SOC), the diamond-shaped nodal line will generically be gapped (also hybridized with other bands). However, as the SOC gaps are small and more than $200 \text{meV}$ below the Fermi energy, our simple tight-binding model still gives a good description of the spin-orbital textures of the Fermi surface that are relevant for superconductivity and transport measurements. The SOC effect on the nodal line along $X$-$M$ also vanishes to leading order. Therefore we neglect SOC in our tight-binding model, and the model band Hamiltonian is 
\begin{equation} \label{eq:TBModel_TBBandTot}
  \mathcal{H} = \mathcal{H}_{nn} + \mathcal{H}_{2nn}  + \mathcal{H}_{\mu}
\end{equation}

We now assume the mean field approximation applies to $\ch{LaCuSb_2}$ and discuss the superconducting pairings according to Bogoliubov-de Gennes effective Hamiltonian. We write the generic superconducting paring terms for Cooper pair with zero center-of-mass momentum as
\begin{equation}\label{eq:SC_GenericSCPairing}
\begin{split}
\mathcal{H}_{\Delta} 
=&\sum_{\mathbf{k}} 
\psi_{-\mathbf{k}, \sigma, r, p_{\alpha}}^\dagger 
\hat{\Delta}_{\sigma, r, p_{\alpha}; \sigma', r', p_{\beta}}(\mathbf{k})
\psi_{\mathbf{k}, \sigma', r', p_{\beta}}^\dagger
+ \text{h.c.},
\end{split}
\end{equation}
where $\hat{\Delta}^T(-\mathbf{k}) = -\hat{\Delta}(\mathbf{k})$ is required by the anticommutation relation for the fermionic operators. The transpose $T$ acts on all the indices including spins, sublattices, and orbitals. We choose the basis of matrices
\begin{equation}\label{eq:SC_TensorProdPauli}
\begin{split}
\tau_\mu^{(\sigma)} \otimes \tau_{\nu}^{(\alpha)} \otimes \tau_{\rho}^{(r)},
\end{split}
\end{equation}
Therefore, there are $28$ possible terms. From current experimental results, the superconductivity is  consistent with singlet pairing, so we can fix the spin part to be $\mathrm{i} \tau_y^{(\sigma)}$. Transforming Eq.~\ref{eq:SC_GenericSCPairing} back to real space, we assume that the leading pairing terms come from the local on-site attractive interactions, which further fixes pairing in the sublattice space to be $\tau_0^{(r)}$.
There remain three possibilities for orbital degrees of freedom i.e., $\tau_0^{(\alpha)}$, $\tau_1^{(\alpha)}$, $\tau_3^{(\alpha)}$.
Therefore, we have
\begin{equation}\label{eq:SC_DeltaParam}
\begin{split}
\hat{\Delta} =& \Delta_1 \mathrm{i}\tau_y^{(\sigma)} \otimes \tau_{0}^{(\alpha)} \otimes \tau_{0}^{(r)}\\
&+ \Delta_2 \mathrm{i}\tau_y^{(\sigma)} \otimes \tau_{1}^{(\alpha)} \otimes \tau_{0}^{(r)}\\
&+ \Delta_3 \mathrm{i}\tau_y^{(\sigma)} \otimes \tau_{3}^{(\alpha)} \otimes \tau_{0}^{(r)}.
\end{split}
\end{equation}
Although the pairing amplitudes $\Delta_i$ are allowed to take generic complex numbers, an over-all $U(1)$ phase is chosen by spontaneous symmetry breaking of the superconductivity. Furthermore, notice that if $\Delta_1/\Delta_3$ is not purely imaginary, the pairing amplitudes for $p_x$- and $p_y$-orbitals are anisotropic, which has not been evidenced from the experiments. The resulting BdG spectrum is shown in Extended Fig.~\ref{fig:TightBinding}g,h, which highlight the anisotropic gap that results from the theory, with typical values of pairing amplitudes $\Delta_1=\Delta_2 = 20\text{meV}$. The pairing amplitudes are set large compared to the experimental value determined from $T_c$ for illustrative purposes. Zooming in on the spectrum near $X$, extended Fig.~\ref{fig:TightBinding}h shows that although the pairing amplitudes are isotropic, very anisotropic gaps are induced on the Fermi surfaces. This anisotropy originates from the spin-orbital textures of the Dirac nodal lines. While the detailed profiles of the gap size depends on various combinations of the pairing amplitude $\Delta_i$, the anisotropy induced by the spin-orbit texture is generic.

\section{Gap fitting models} \label{ssec:SI_GapFit}
The complicated structure of the gap function discussed in the tight-binding analysis, along with the many free parameters, makes microscopic fitting untenable. We discuss below how we can try to faithfully represent the gap function using fewer parameters. To start, in the alpha model, the gaps $ \Delta_i(0)/k_B T_{ci} \equiv 1.764\cdot \alpha_i$ are taken as variables that can differ from the nominal BCS value $\alpha_0 = 1$. In principle $T_c$ can also be different among the gaps. The specific heat in the superconducting state is obtained from the entropy \citesec{Bouquet},
 \begin{equation}
     \frac{S_i}{\gamma_i} = -\frac{6}{\pi^2 k_B} \int_0^\infty [f\ln f + (1-f)\ln(1-f) ] d\epsilon 
 \end{equation}
where $f = 1/(\exp(\beta E) + 1)$, with $\beta= (k_B T)^{-1}$, and $E = \sqrt{\epsilon^2 + \Delta_i^2 (T)}$. The gap can be modeled approximately with the BCS gap modified by the $\alpha$-model \citesec{Carrington}
 \begin{equation}
     \Delta_i(T) = \alpha \cdot 1.764 k_B T_{ci} \tanh \{ 1.82 [1.018 (T_c/T-1)]^{0.51} \} 
 \end{equation} 
The specific heat contribution from each gap is 
 \begin{equation}
     \frac{c_i}{\gamma_i T} = \dv{(S_i/\gamma_i)}{T}\bigg|_T
 \end{equation}
and the total specific heat is 
 \begin{equation}
     \frac{c}{\gamma_n T} = \frac{1}{\gamma_n T} \sum_i c_i = \sum_i n_i \cdot \frac{c_i}{\gamma_i T}
 \end{equation}
with $\gamma_i/\gamma_n = n_i$ and $\sum_i n_i = 1$. 

The model should also be consistent with the superfluid density extracted from $\mu$SR. In the dirty limit, the superfluid density is 
 \begin{equation} \label{eq:dirty_rho}
     \rho_\mu(T) \equiv \frac{\lambda^2_\mu(0)}{\lambda^2_\mu(T)} = \frac{\Delta_\mu(T)}{\Delta_\mu(0)} \tanh\left( \frac{\Delta_\mu(T)}{2k_B T} \right)
 \end{equation}
The measured superfluid density would then be related to the two separate contributions by 
 $$ \rho(T) = \sum_\mu \gamma_\mu \rho_\mu (T) $$
with $\sum_\mu \gamma_\mu = \gamma_1 + (1-\gamma_1) = 1$. For two bands the $\gamma_1$ parameter depends on the Fermi velocities of the bands as follows
 \begin{equation} \label{eq:gamma_parameter}
     \gamma_1 = \frac{n_1 v_1^2}{n_1 v_1^2 + n_2 v_2^2}
 \end{equation} 
While in our BdG model the gap can be anisotropic, we use a single parameter for the gap function (in each band) as an effective gap size.

The alpha-model explicitly assumes that there is no inter-band pairing interaction, however such an interaction cannot be ruled out. Complementary to the alpha-model is the self-consistent Eilenberger two-band model \cite{Prozorov2} which accounts for inter-band pairing. To calculate the relevant thermodynamic quantities for the superconducting state, we write $\delta_\mu = \Delta_\mu/2\pi T$ for the bands $\mu=1,2$, and we self-consistently solve the coupled equations 
 \begin{gather} 
     \delta_\nu = \sum_\mu n_\mu \lambda_{\nu\mu} \delta_\mu \cdot \left( \widetilde{\lambda}^{-1} + \ln \frac{T_c}{T} - A_\mu \right) \\
     A_\mu = \sum_{n=0}^\infty \left[ \frac{1}{n+1/2} - \frac{1}{\sqrt{\delta_\mu^2 + (n+1/2)^2}} \right] 
 \end{gather}
where 
 \begin{equation}
     \widetilde{\lambda} = \frac{2 n_1 n_2 (\lambda_{11} \lambda_{22} - \lambda_{12}^2)}{n_1 \lambda_{11} + n_2 \lambda_{22} - \sqrt{ (n_1 \lambda_{11} - n_2 \lambda_{22})^2 + 4n_1 n_2\lambda_{12}^2 }} 
 \end{equation}
and $\lambda_{\nu\mu} = N(0) V(\nu,\mu)$ are dimensionless effective interaction coefficients. In the clean limit
 \begin{equation}
     \rho_\mu(T) = \sum_{n=0}^\infty \frac{\delta_\mu^2}{[\delta_\mu^2 + (n+1/2)^2]^{3/2}}
 \end{equation}

We have modeled the specific heat data using various models: (1) single BCS-like gap, (2) two-gap $\alpha$-model with different $T_c$; (3) and two-gap Eilenberger model. The specific heat and superfluid density were simultaneously refined by minimizing $\chi^2 = \chi^2_\mathrm{c} + \chi^2_{\rho}$. The results of the fit are shown in Table~\ref{tab:AlphaModels}. We emphasize that the fits may suffer from the slight deviation in the data from entropy balance, which requires $\Delta S(T_c)/\gamma_n T_c = 1$. Furthermore, due to the many parameters in the tight-binding and BdG models, we cannot at present select a specific model for \ch{LaCuSb2}. However, it is worth noting that  Model 3 comes the closest to matching the superfluid density in the dirty limit, primarily due to the influence of one gap (as $\gamma_1 \approx 1)$. Simultaneously the curvature of the second gap comes closer to representing the specific heat drop near $T^*$.

\begin{figure}
    \centering
    \includegraphics[width=\linewidth]{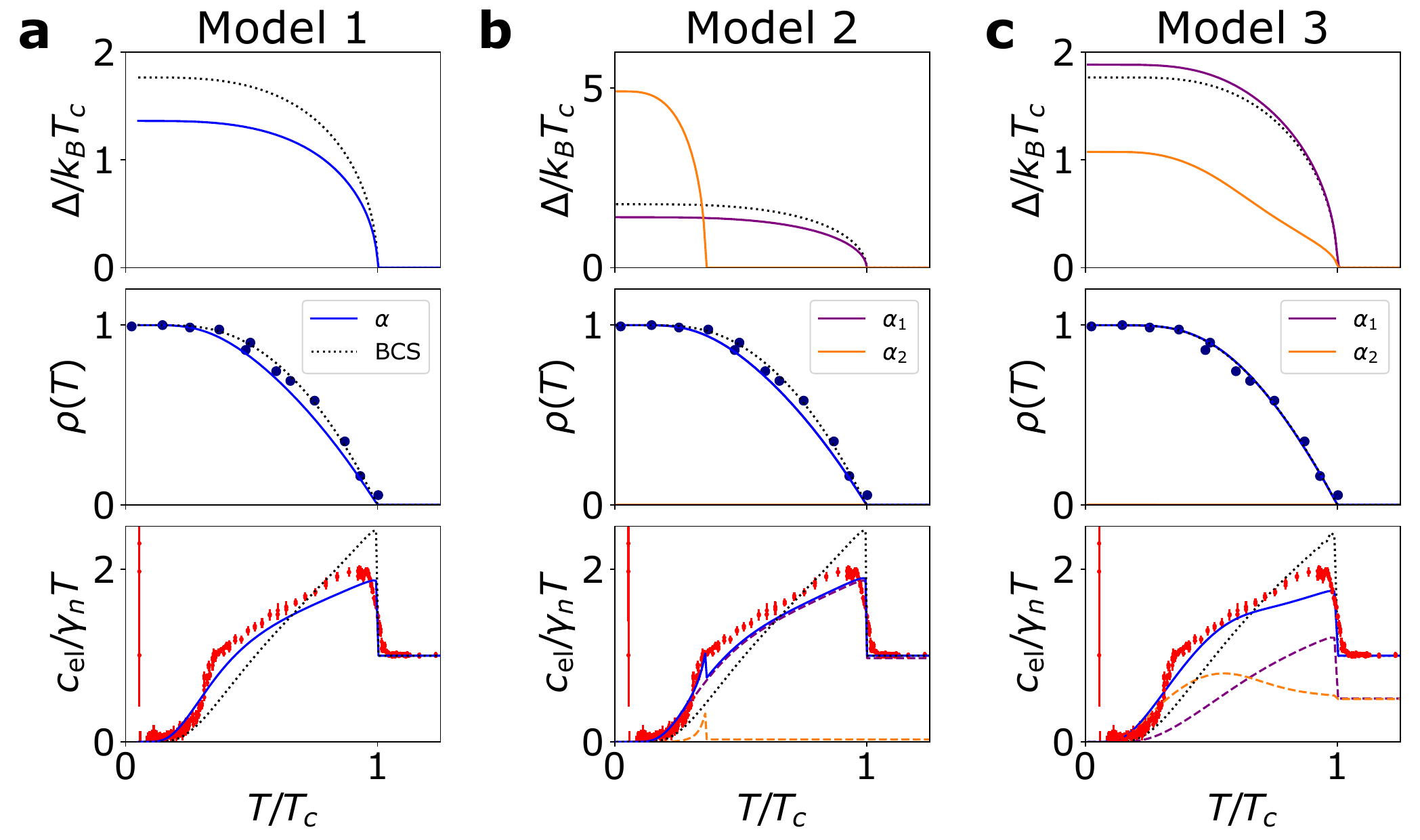}
    \caption{ \edfigurelabel{fig:LCS_Models} Various simplified models of superconducting gaps in \ch{LaCuSb2} as described in the text, along with fits to the superfluid density measured by transverse field $\mu$SR and to the electronic specific heat. Model 1 corresponds to a single gap $\alpha$-model. Model 2 corresponds to a two-gap $\alpha$-model with two independent superconducting transition temperatures. Model 3 corresponds to the Eilenberger two-band model with bands opening at the same $T_c$.} 
\end{figure} 

\section{Phase diagram} \label{SI_phasediagram} 
From the specific heat, magnetization, resistivity, and $\mu$SR data, we deduced the field-temperature phase diagram and estimated the critical field. We show the complete phase diagram for magnetic fields applied along the $a$- and $c$-axis in Extended Fig.~\ref{fig:phase_diagram}. Using fits to the critical fields as a function of temperature, we find that $H_{c1}(0) = 32(1)$ Oe and $H_{c2}(0) = 172(6)$ Oe for the Type-II superconducting state using magnetization data, whereas $H_{c}(0) \approx 65.1(2)$ Oe in the Type-I superconducting state using $\mu$SR data. 

The thermodynamic critical field $H_c$ as deduced in a Type-II superconductor can be estimated from $H_c \approx \sqrt{H_{c1} H_{c2}}$. To verify if this is the case, we plotted $\sqrt{H_{c1} H_{c2}}$ as a function of temperature using the magnetization data in the Type-II superconducting state. As seen in Extended Fig.~\ref{fig:phase_diagram}, the resulting data are quite comparable to the critical fields derived from the Type-I superconducting state, which affirms that both the anisotropy and low critical fields are intrinsic to \ch{LaCuSb2} and its particular Dirac band structure.

\begin{figure}
    \centering
    \includegraphics[width=\linewidth]{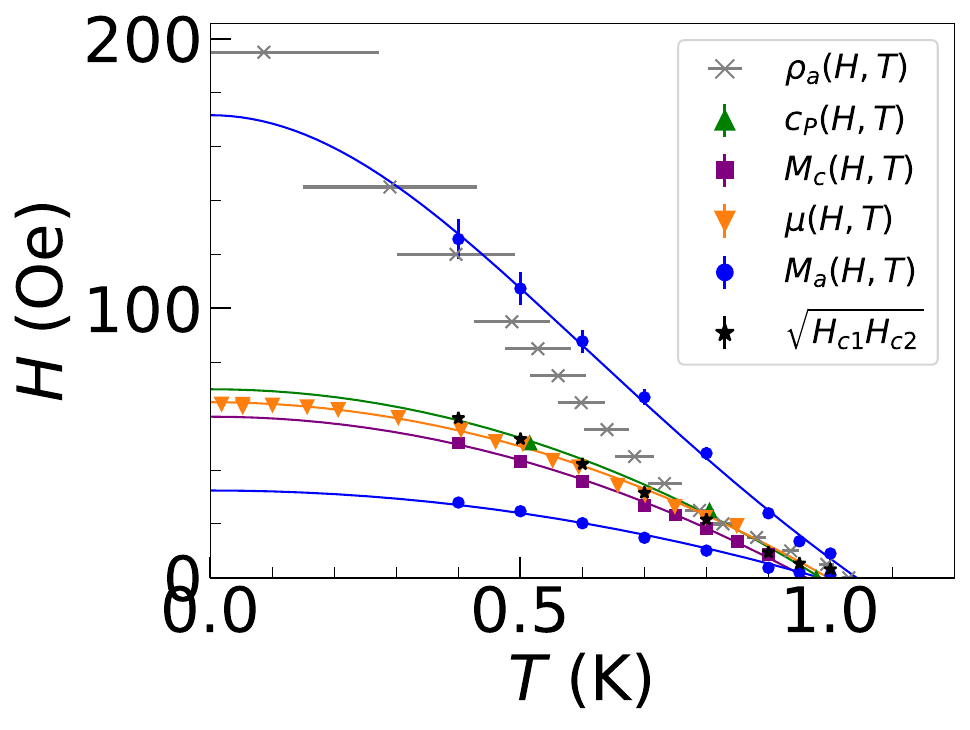}
    \caption{ \edfigurelabel{fig:phase_diagram} Phase diagram for \ch{LaCuSb2} constructed from magnetization $M_a$ and $M_c$, specific heat $c_p$, muon spin rotation $\mu$ for fields along the $c$-axis, and resistivity data $\rho_a$ for in-plane currents and out-of-plane fields. Points are the critical points deduced from the data as indicated throughout the SI, and solid lines are fits to Eq.(~\ref{eq:critical_fields}).} 
\end{figure}



\begin{table}[ht!]
\footnotesize
\begin{tabular}{ |p{0.28\textwidth} | c | c | }
\hline {\bf Quantity} &  {\bf Equation} & {\bf Computed value} \\ \hline\hline 
     transition $T_c$ &   & 0.98(2) K   \\ \hline
     Sommerfeld constant $\gamma_n$ &   & 4.78(1) mJ/mol-K$^2$  \\ \hline 
     Resistivity plateau $\rho_{0a}$ &  & $1.883(15)$ $\mu\Omega$-cm \\ \hline 
     carrier density $n$ &    & $  4.16(3)\times 10^{28}$ m$^{-3}$  \\ \hline 
     BCS gap $\Delta(0)$ & $1.764 k_B T_c$ & 0.15 meV \\ \hline
    Fermi momentum $k_F$ &  $ (3 \pi^2 n)^{1/3}$  & $1.072(3) \times 10^{10}$ m$^{-1}$ \\ \hline 
       specific heat effective mass $m^*$ & $\dfrac{(\hbar k_F)^2 \gamma_n}{\tfrac{1}{2} \pi^2 V_0 Rk_B n}$                                          & $  1.44(1) m_e$ \\ \hline 
   Pauli limit (mks) $H_p$ & $1.84 T_c$  & 1.80(4) T   \\ \hline 
   BCS critical field (cgs) $H_c(0)$ & $1.764 \sqrt{\dfrac{6}{\pi}} \cdot \gamma_{nV}^{1/2} T_c$  & 68(1) Oe  \\ \hline 
       Fermi velocity $v_F$ &  $\dfrac{\hbar k_F}{m^*}$                 & $8.59(7)\times 10^5$ m/s  \\ \hline 
      scattering time $\tau$ &  $\dfrac{m^*}{ne^2 \rho_0}$              & $6.55(8) \times 10^{-14}$ s  \\ \hline 
       mean free path $\ell$ &  $v_F \tau$                                 & $5.62(8)\times 10^{-8}$ m  \\ \hline 
   coherence length $\xi$  &  $0.18 \dfrac{\hbar v_F}{k_B T_c}$        & $1.21(2)\times 10^{-6}$ m  \\ \hline 
    penetration depth $\lambda_L $ &  $\sqrt{ \dfrac{m^*}{\mu_0 ne^2}}$  & $3.13(3)\times 10^{-8}$ m \\ \hline 
    penetration depth $\lambda(0)$, $\mu$SR & $(\lambda_{a} \lambda_{c})^{1/2}$  & $4.08(2) \times 10^{-7}$ m  \\ \hline
    clean/dirty limit &  $\xi/\ell$                                        & $ $ 21.5(5) (dirty) \\  \hline 
 GL parameter $\kappa_c$ (clean) &  $\lambda_L/\xi$  & 0.0249(6) \\  \hline 
 GL parameter $\kappa_c$ (dirty) &  $0.715 \lambda_L/\ell $ & 0.398(7) \\  \hline   
 GL parameter $\kappa_a$ &  $\dfrac{H_{c2}}{\sqrt{2} H_c}$ & 2.03(8) \\  \hline
\end{tabular}
\caption{ \edfigurelabel{tab:SC_quantities} Parameters estimated from experimental results, as well as under the assumption of an isotropic free-electron model and Ginzburg-Landau theory.}
\end{table}

\begin{table}[ht!]
\begin{tabular}{c|c|c||c|c} 
  $\alpha$ & Model 1  & Model 2 & $\gamma$    & Model 3     \\ \hline 
$T_{c1}$   & 0.978    & 0.978  & $T_{c}$        & 0.978    \\
$n_1$      & 1        & 0.97   & $n_1$          & 0.502(8) \\
$\gamma_1$ & 1        & 1.00   & $\gamma_1$     & 0.99(3)  \\ 
$\alpha_1$ & 0.771(8) & 0.794  & $\lambda_{11}$ & 0.91(3)  \\
$\alpha_2$ &          & 2.78   & $\lambda_{22}$ & 0.68(12) \\
$T_{c2}$   &          & 0.361  & $\lambda_{12}$ & 0.06(3)  
\end{tabular} 
\caption{ \edfigurelabel{tab:AlphaModels} Fitted parameters using (1) one-gap alpha model; (2) two-gap alpha model assuming two different $T_c$; and (3) the Eilenberger self-consistent two-band model.} 
\end{table}

\pagebreak 

\noindent{\bfseries Other references}\setlength{\parskip}{12pt}%
\bibliographystylesec{naturemag}

\end{document}